\documentclass[%
superscriptaddress,
amsmath,amssymb,
 aps,prd,
 twocolumn,
 tightenlines,
 nofootinbib
]{revtex4-1}

\usepackage{graphicx,url,amssymb,amsmath,longtable,rotating,color,units,wasysym,subfigure,epsfig,hyperref,epstopdf}
\usepackage[dvipsnames]{xcolor}
\usepackage{enumitem}
\usepackage[title]{appendix}

\usepackage{graphicx}
\usepackage{dcolumn}
\usepackage{bm}
\usepackage{url,hyperref}
\usepackage{array,multirow}
\usepackage{color} 
\usepackage{aas_macros} 
\usepackage{ifthen}
\usepackage{etoolbox}
\usepackage[export]{adjustbox}

\usepackage{siunitx} 
\usepackage{acronym,cleveref}
\usepackage{soul}
\definecolor{darkgreen}{rgb}{0.0, 0.5, 0.0}

\newcommand{\Ogw}{\Omega_\text{GW}}

\newcommand{\fref}{f_\text{ref}}

\newtoggle{fullauthorlist}
\toggletrue{fullauthorlist}

\newtoggle{endauthorlist}
\toggletrue{endauthorlist}

\usepackage{xr}

\acrodef{LIGO}{Laser Interferometer Gravitational-wave Observatory}
\acrodef{aLIGO}{Advanced LIGO}
\acrodef{GW}{gravitational waves}
\acrodef{CGB}{cosmological gravitational-wave background}
\acrodef{SNR}{signal-to-noise ratio}

\begin{document}

\title{Search for anisotropic gravitational-wave backgrounds using data from Advanced LIGO and Advanced Virgo's first three observing runs}
\date{\today}

\iftoggle{endauthorlist}{
  %
  %
  \let\mymaketitle\maketitle
  \let\myauthor\author
  \let\myaffiliation\affiliation
  \author{The LIGO Scientific Collaboration}
  \author{The Virgo Collaboration}
  \author{The KAGRA Collaboration}
  \email{Full author list given at the end of the article.}
\noaffiliation
}{
  %
  %
  \iftoggle{fullauthorlist}{
  }{
    \author{The LIGO Scientific Collaboration}
    \affiliation{LSC}
    \author{The Virgo Collaboration}
    \affiliation{Virgo}
    \author{The KAGRA collaboration}
    \affiliation{KAGRA}
  }
}

\begin{abstract}
We report results from searches for anisotropic stochastic gravitational-wave backgrounds using data from the first 
three observing runs of the Advanced 
LIGO and Advanced Virgo detectors. For the first time, we include Virgo data in our analysis and run our search with a new efficient pipeline called {\tt PyStoch} on data folded over one sidereal day. We use gravitational-wave radiometry (broadband and narrow band) to produce sky maps of stochastic gravitational-wave backgrounds and to search for gravitational waves from point sources. A spherical harmonic decomposition method is employed to look for gravitational-wave emission from spatially-extended sources. Neither technique found evidence of gravitational-wave signals.
Hence we derive 95\% confidence-level upper limit sky maps on the gravitational-wave energy flux from broadband point sources, ranging from $F_{\alpha, \Theta} < {\rm (0.013 - 7.6)} \times 10^{-8} {\rm erg \, cm^{-2} \, s^{-1} \, Hz^{-1}},$ and on the (normalized) gravitational-wave
energy density spectrum from extended sources, ranging from $\Omega_{\alpha, \Theta} < {\rm (0.57 - 9.3)} \times 10^{-9} \, {\rm sr^{-1}}$, depending on direction ($\Theta$) and spectral index ($\alpha$). These limits improve upon previous limits by factors of $2.9 - 3.5$. We also set 95\% confidence level upper limits on the frequency-dependent strain amplitudes of quasimonochromatic gravitational waves coming from three interesting targets, Scorpius X-1, SN 1987A and
the Galactic Center, with best upper limits range from $h_0 < {\rm (1.7-2.1)} \times 10^{-25},$ a factor
of $\geq 2.0$ improvement compared to previous stochastic radiometer searches.
\end{abstract}

\maketitle

\section{Introduction} \label{sec:introductions}

The stochastic gravitational-wave background (GWB) is composed of a combination of gravitational-wave signals from many unresolved sources \cite{maggiore, Christensen:2018iqi}. A major contribution is expected to be of astrophysical origin, i.e., produced by the superposition of gravitational-wave signals from unresolved individual sources such as binary black hole and neutron star mergers \cite{tania, Regimbau_2011,Banagiri:2020kqd,Payne:2020pmc,Stiskalek:2020wbj}, supernovae \cite{2009MNRAS.398..293M,2010MNRAS.409L.132Z,PhysRevD.72.084001,PhysRevD.73.104024,marocmodi}, or depleting boson clouds around black holes \cite{Brito:2017wnc,Brito:2017zvb,Fan:2017cfw,Tsukada:2018mbp,palomba2019direct,sun2019search}.
The background may also include signals of
cosmological origin, i.e., produced in the early Universe during an inflationary epoch\cite{1994PhRvD..50.1157B,1979JETPL..30..682S,2007PhRvL..99v1301E,2012PhRvD..85b3525B,2012PhRvD..85b3534C,Lopez:2013mqa,1997PhRvD..55..435T,2006JCAP...04..010E,axion_inflation}, or as a direct result of phase transitions \cite{vonHarling:2019gme,Dev:2016feu,Marzola:2017jzl}, primordial black hole mergers \cite{MandicEA_2016,SasakiEA_2016,Wang:2016ana,Miller:2020kmv}, or
other associated phenomena \cite{Caprini_2018}.
Different models could, in principle, be distinguished by characteristic features in the angular distribution \cite{contaldi_aniso,Jenkins:2018kxc, Jenkins:2019uzp,Jenkins:2019nks,Bertacca:2019fnt,cusin_pitrou_analytic_expression_angular_power,Cusin:2017mjm,Cusin:2018rsq,Cusin:2019jpv,Pitrou:2019rjz,Canas-Herrera:2019npr,Geller:2018mwu}. For example, cosmic strings have an angular power spectrum which is sharply peaked at small multipoles \cite{Jenkins:2018lvb,Jenkins:2018uac}, while neutron stars in our Galaxy would trace out the Galactic plane \cite{talukder2014measuring,mazumder2014astrophysical}. 
In this paper we search for an anisotropic GWB using data from the Advanced LIGO~\citep{2015CQGra..32g4001L} and Advanced Virgo~\cite{TheVirgo:2014hva} gravitational-wave detectors.  This is the first time we have included data from Virgo in a search for an anisotropic GWB \cite{abbott2017directional,abbott2019directional}. 

The three analyses presented in this paper rely on cross-correlation techniques~\cite{romano2017detection}, which have been employed extensively on gravitational-wave data in the past, and are referred to as the broadband radiometer analysis (BBR) ~\cite{radio_method,radiometer}, the spherical harmonic decomposition (SHD)~\cite{sph_methods,sph_results}, and the narrow band radiometer analysis (NBR)~\cite{Ballmer2006LIGOIO}. The BBR analysis targets a small number of resolvable, persistent point sources emitting gravitational waves over a wide frequency band. The SHD analysis reconstructs the harmonic coefficients of the gravitational-wave power on the sky, and can identify extended sources with smooth frequency spectra. Finally, the NBR analysis studies frequency spectra from three astrophysically relevant sky locations: Scorpius X-1 \cite{abbott2017upper,abbott2017search}, Supernova 1987A \cite{Sun:2016idj,chung2011designing}, and the Galactic Center \cite{aasi2013directed,dergachev2019loosely}. 
Resolvable point sources in the sky are not expected to follow an isotropic distribution \cite{gwtc-1}, underscoring the importance of analysis techniques that can deal with anisotropic backgrounds. 

For the first time, we employ data folding, a technique that takes advantage of the temporal symmetry inherent to Earth's rotation, to combine the data from an entire observation run into one sidereal day, greatly reducing the computational cost of this search \cite{ain2015fast}. Furthermore, we have employed the python based pipeline {\tt PyStoch} \cite{ain2018very} to perform the analyses on folded data reported in this paper. 

We do not find evidence for gravitational waves in any of the three analyses and hence set direction-dependent upper limits on the gravitational-wave emission.  Though stringent upper limits on the anisotropic GWB have been obtained in the past \cite{abbott2017directional,abbott2019directional,Renzini:2019vmt}, our new constraints improve upon existing limits by a factor of $\geq 2.0$. 

This paper is structured as follows: Sec \ref{sec:methods} presents the GWB model adopted in our analyses, and the search methods used. Section \ref{sec:data} describes the datasets used in the searches and briefly explains the data processing. Results from all three analyses are presented in Section~\ref{sec:results}. Finally, we conclude with the interpretation of our results in Sec~\ref{sec:conclusions}.

\section{Methods}\label{sec:methods}

The goal of the anisotropic GWB search is to estimate gravitational-wave power as a function of sky direction and model its spatial distribution. 
The analyses presented in this paper use the methods described in \cite{sph_methods, romano2017detection, Mitra_2008_radio_method2}. Assuming an unpolarized, Gaussian and stationary GWB, the quadratic expectation value of the gravitational-wave strain distribution $h_A(f,\Theta)$ across different sky directions and frequencies can be written as
\begin{equation}
    \langle h_A^*(f, \Theta) \, h_{A'}(f', \Theta') \rangle = \frac{1}{4} \mathcal{P}(f, \Theta) \, \delta_{AA'} \, \delta(f-f') \, \delta(\Theta, \Theta') \, ,
\end{equation}
where $A$ represents gravitational-wave polarization, asterisk ($^*$) denotes 
the complex conjugate and $\mathcal{P}(f, \Theta)$ characterizes the gravitational-wave strain power as a function of frequency $f$ and direction $\Theta$. As in previous searches \cite{abbott2017directional,abbott2019directional} and suggested in the literature \cite{allen-ottewill, romano2017detection}, we factorize $\mathcal{P}(f, \Theta)$ into frequency and sky-direction dependent components,
\begin{equation}
 \mathcal{P}(f, \Theta) = H(f) \, \mathcal{P}(\Theta),
\end{equation}
where $H(f)$ describes the spectral shape and $\mathcal{P}(\Theta)$ denotes the angular distribution of gravitational-wave power. In our analyses we model the spectral dependence $H(f)$ as a power-law given by
 \begin{equation}
 H(f) =\left(\frac{f}{f_{\rm ref}}\right)^{\alpha -3}, 
 \end{equation}
where $\alpha$ is the spectral index and $f_{\rm ref}$ is a reference frequency set to 25 Hz, as in past searches \cite{abbott2017directional,abbott2019directional}. We consider three values of $\alpha$ corresponding to different GWB physical models: $\alpha=0$, consistent with many cosmological models, such as slow roll inflation and cosmic strings, in the observed frequency band~\cite{Caprini_2018}; $\alpha=2/3$, compatible with an astrophysical background dominated by compact binary inspirals \cite{sesana2008stochastic}; and $\alpha=3$, indicating a generic flat strain spectrum \cite{gw150914_stoch}.

We define the cross-correlation spectra from two detectors $(I,J)$ evaluated at time $t$ and frequency $f$ as \cite{sph_methods, romano2017detection}
\begin{equation}
    C_{IJ}(t;f) = \frac{2}{\tau} \tilde{s}^*_I(t; f) \, \tilde{s}^*_J(t; f) ,
\end{equation}
where $\tilde{s}(t;f)$ is the short-time Fourier transform of time segment $s(t)$ of duration $\tau$.
As shown in \cite{sph_methods}, the quadratic expectation value of gravitational-wave strain can be related to the above  cross-correlation spectra $C_{IJ}(t;f)$ by
\begin{equation}
    \label{eq:cross_spectrum}
    \langle C_{IJ}(t; f) \rangle = H(f) \int_{S^2} d \Theta \, \gamma_{IJ}(t; \Theta, f) , \mathcal{P}(\Theta)
\end{equation}
where $\gamma_{IJ}(t; \Theta,f)$ is a geometric function which encodes the combined response of a detector pair to gravitational waves \cite{sph_methods}. The right-hand side of Eq. \eqref{eq:cross_spectrum} can be rewritten in terms 
of a set of basis functions, labeled by $\mu$, on the two-sphere ($S^2$) as
\begin{equation}
    \langle C_{IJ}(t; f) \rangle = H(f) \, \gamma_\mu \, \mathcal{P}_\mu ,
\end{equation}
where the summation (or integration) over $\mu$ is understood. For the SHD analysis reported in this paper, we employ the spherical harmonics basis $\mu\rightarrow \ell m$ and for the BBR and NBR analyses we choose the pixel basis $\mu \rightarrow \Theta$.
In the weak signal limit, the covariance matrix of the cross-spectra $C_{IJ}(t;f)$ is given by \cite{romano2017detection}
\begin{equation}
    N_{ft,f't'} = \delta_{tt'} \, \delta_{ff'} \, P_I(t;f)  P_J(t;f),
\end{equation}
where $P_I$ is the one-sided power spectrum of the data from detector $I$.

Assuming a fiducial model for the signal spectral shape $H(f)$ and further assuming the detector noise spectra are well estimated, the likelihood function relating $C_{IJ}(t;f)$ and $\mathcal{P}_\mu$ can be written as,
\begin{align}
    p(C_{IJ}(t; f)|\mathcal{P}_\mu) \propto &  \, {\rm exp} \Big( [C_{IJ}(t;f)- H(f) \, \gamma_\mu \, \mathcal{P}_\mu]^* \nonumber \Big. \\ & \Big. N^{-1}_{ft,f't'}  [C_{IJ}(t;f)- H(f) \, \gamma_\mu \, \mathcal{P}_\mu] \Big).
\end{align}

  Maximizing the above likelihood function for $\mathcal{P}_\mu$ we get \cite{sph_methods}
\begin{equation}
    \label{eq:cleanmap}
    \mathcal{P}_\mu = (\Gamma^{-1})^{IJ}_{\mu \nu} \, X_{\nu}^{IJ}  ,
\end{equation}
 where
 \small
 \begin{eqnarray}
     \label{eq:dirtymap}
     X_{\nu}^{IJ} = \sum_t \sum_f (\gamma_{IJ})_{\nu}^*(t; f) \frac{H(f)}{P_I(t; f) P_J(t; f)} C_{IJ}(t; f), \\
     \label{eq:fisher}
     \Gamma_{\mu \nu}^{IJ} = \sum_t \sum_f (\gamma_{IJ})_{\mu}^*(t; f) \frac{H^2(f)}{P_I(t; f) P_J(t; f)} (\gamma_{IJ})_{\nu}(t; f).
 \end{eqnarray}
 \normalsize
 The vector $X_\nu$, often referred to as the ``dirty map", is a convolution of the gravitational-wave power sky map with the directional response function of a given baseline $IJ$, and $\Gamma_{\mu \nu}$ is called the Fisher information matrix.
 For a network of detectors with multiple baselines, the combined $X_{\nu}$ and $\Gamma_{\mu \nu}$ can be obtained by summing over all baseline contributions as
 \begin{eqnarray}
    \label{eq:combine_map}
    X_\nu &=& \sum_I \sum_{J>I} X^{IJ}_{\nu}, \\
    \label{eq:combine_Fisher}
    \Gamma_{\mu\nu}& =& \sum_I \sum_{J>I} \Gamma^{IJ}_{\mu\nu}.
\end{eqnarray}

Using the above Fisher matrix and dirty map, we estimate the GWB power $\mathcal{P}_\mu$, referred to as the ``clean map'':
\begin{equation}
    \label{eq:cleanmap_regularized}
    \hat{\cal P}_\mu = \sum_\nu\,\left(\Gamma_{\rm R}^{-1}\right)_{\mu\nu} X_{\nu}\,,
\end{equation}
which requires inverting the Fisher matrix $\Gamma_{\mu\nu}$. However, the Fisher matrix tends to be singular as the detector pairs are insensitive to certain sky directions or $\ell m$ modes, and hence a full inversion cannot be performed. Therefore we use a regularized pseudoinverse (labeled by the subscript `R' above) to obtain clean maps.
We note here that $[\left(\Gamma_{\rm R}^{-1}\right)_{\mu\mu}]^{1/2}$ is used as the uncertainty estimate (standard deviation) of $\hat{\cal P}_\mu$.

Different regularization techniques are employed in each analysis based on the signal model assumed~\cite{abbott2017directional}. 
For the BBR search we assume that the gravitational-wave power is confined to a single pixel and there is no signal covariance between neighboring pixels; hence, the inversion of the Fisher matrix reduces to the inversion of its diagonal. 
However, because of the detector response function, neighboring pixels are indeed correlated and hence the BBR results are valid only for a signal model in which we expect a small number of well-separated gravitational-wave point sources. 

On the other hand, the SHD analysis uses both the diagonal and off-diagonal elements of the Fisher matrix and as in past searches sets the smallest 1/3 of the eigenvalues to infinity and also uses a finite maximum value of $\ell$~\cite{abbott2017directional,sph_methods,sph_results}. The choice of $1/3$ is based on the recovery of simulated injections carried out in reference \cite{sph_methods}. This analysis is therefore well suited for identifying extended sources on the sky, but not pointlike sources which require all the $\ell$ modes with $\ell \to \infty$. SHD analyses of the previous two LIGO/Virgo observing runs chose the maximum $\ell$ value $\ell_{\rm max}$ based on 
the diffraction-limited angular resolution $\theta$ on the sky.
This is determined by the distance $D$ between detectors and the most sensitive frequency $f$ in the analysis band~\cite{abbott2017directional}:
\begin{equation}
    \theta = \frac{c}{2Df} \hspace{1cm} \ell_{\rm max}=\frac{\pi}{\theta}.
\end{equation}
As in the previous directional searches, this method gives $\ell_{\rm max}$ values of 3, 4, and 16 for the spectral indices $\alpha$ of 0, 2/3, and 3, respectively, for the Hanford-Livingston baseline. The most sensitive frequency in the analysis changes with $\alpha$ and hence we get different $\ell_{\rm max}$ for different $\alpha$. The baseline sensitivity ($\propto 1/[P_I P_J]$) appearing in Eqs.~\eqref{eq:dirtymap} and \eqref{eq:fisher} acts as a weighting factor multiplying $\gamma_{IJ}^{\ell m}(t;f)$, and hence, the cutoff on $\ell$ also depends on the baseline's sensitivity among the network. Since the LIGO detectors are more sensitive than the Virgo detector, $\ell_{\rm max}$ values are largely determined by the Hanford-Livingston baseline. 
Therefore, in this search, we make the same choices for $\ell_{\rm max}$ for all baselines in the Hanford-Livingston-Virgo network. 

We note that, as described in \cite{Renzini:2018nee,Renzini:2019vmt,Renzini:2018vkx,Suresh:2020khz}, one could also start in a pixel basis and transform the resultant pixel-based maps into spherical harmonic coefficients. Sampling the full pixel space accounts for the correlations between small and large angular scales induced by the noncompactness of the sky response (for details see \cite{Renzini:2018vkx}). 
  
In the SHD analysis we calculate $\hat{\mathcal{P}}_{\ell m}$ in the spherical harmonics basis and express the final result in terms of $\hat{C}_{\ell}$, a measure of squared angular power in mode $\ell$, which is given by \cite{sph_methods}
\begin{equation}
    \hat C_{\ell} = \left(\frac{2\pi^2 \fref^3}{3H_0^2}\right)^2\frac{1}{1 + 2 \ell} \sum_{m=-\ell}^{\ell} \left[|\hat{\mathcal P}_{\ell m}|^2 - (\Gamma_{\rm R}^{-1})_{\ell m,\ell m}\right],
\end{equation}
where $H_0$ is the Hubble constant taken to be $H_0 = \unit[67.9]{km\,s^{-1}\,Mpc^{-1}}$ \cite{Planck_2015}.
$\hat{C}_{\ell}$ has units of $\mathrm{sr}^{-2}$ and $\hat{C}_{\ell} = 1$ corresponds to sufficient energy density in mode $\ell$ alone to have a closed universe. 
In addition, we also transform $\hat{\mathcal{P}}_{\ell m}$ to $\hat{\mathcal{P}}_{\Theta}$ 
and produce $\hat{\Omega}_{\alpha, \Theta}$ given by \cite{sph_methods}
\begin{equation}
\hat{\Omega}_{\alpha, \Theta} = \frac{2\pi^2}{3H_0^2} \, f_{\rm ref}^3 \, \hat{\mathcal{P}}_{\alpha,\Theta}\, ,
\end{equation}
which is the gravitational-wave energy density in solid angle $\Theta$ normalized by the critical energy density needed to close the Universe.

In the BBR analysis, we estimate $\mathcal{P}_{\Theta}$ in a pixel basis and report the final result in terms of the gravitational-wave energy flux from solid angle $\Theta$ given by
\begin{equation}
    \label{eq:flux_definition}
    \hat{\mathcal{F}}_{\alpha, \Theta} = \frac{c^3\pi}{4G} \, f_{\rm ref}^2 \hat{\mathcal{P}}_{\alpha, \Theta} \, ,
\end{equation}
where $G$ is the gravitational constant.

In the NBR analysis we measure gravitational-wave strain power $\hat{\mathcal{H}}(f)$ as a function of frequency at specific sky locations by setting $\alpha = 3$ for $H(f)$ and not summing over frequency in Eqs. \eqref{eq:dirtymap} and, \eqref{eq:fisher} i.e., $\hat{\mathcal{H}}(f) = X_{\nu}^{IJ} (f)$.
However, the NBR analysis must consider source-dependent effects when performing a search.
In the case of Scorpius X-1, a low-mass x-ray binary system, gravitational-wave frequencies are expected to be broadened \cite{abbott2017upper} due to the binary motion of the source and the orbital motion of Earth during the observation time \cite{whelan2015model}. To account for these Doppler shifts, we sum the contributions in multiple frequency bins and create optimally-sized \textit{combined bins} at each frequency. For more details of combining frequency bins for Scorpius X-1 see Ref. \cite{abbott2017directional}. In the directions of SN 1987A and the Galactic Center, we combine 3 and 17 frequency bins respectively to account for the spread of an expected monochromatic signal due only to the rotation and orbital motion of the Earth \cite{abbott2017directional}. Since the Galactic Center is at a lower declination, the effect of the Earth's motion becomes significant and hence we combine more frequency bins. 

To perform these three analyses, cross-correlation data from each baseline is folded into one sidereal day by taking advantage of a temporal symmetry of the observations induced by the Earth's daily rotation about its axis. We therefore reduce the computational cost of this search by a factor equal to the total number of days of observation \cite{ain2015fast}.

For the NBR and BBR analyses, the folded data are analyzed by Python-based pipeline, {\tt PyStoch} \cite{ain2018very}, which takes advantage of the compactness of the folded data and the standardization and optimizations of the well-known {\tt HEALPix} (Hierarchical Equal Area isoLatitude Pixelation) Python package \cite{HEALPix} to reduce the computational cost and memory requirements by a factor of a few compared to past analyses. 

\section{Data} \label{sec:data}

For the three analyses, we use data from the third observing run (O3) of Advanced LIGO \citep{2015CQGra..32g4001L} and Advanced Virgo \citep{TheVirgo:2014hva}. The detectors took this data between April 1, 2019 and March 27, 2020, with a one month pause in data collection in October 2019
and had duty factors of $77\%$, $75\%$ and $76\%$ for LIGO Livingston (L), LIGO Hanford (H) and Virgo (V), respectively \cite{detchar2020}.

Similarly to previous analyses \cite{abbott2017directional,abbott2019directional}, we first preprocess the data. The raw time-series strain data are downsampled from 16 kHz  to 4 kHz. Then, the data are divided into 192-second, 50\% overlapping, Hann-windowed segments and filtered through a 16th order Butterworth high-pass filter with a knee frequency of 11 Hz. The 192-second segment duration is chosen so that we can identify narrow spectral features in the data and at the same time not be significantly affected by the changes in the response functions of the detectors due to the Earth's rotation. We then Fourier transform the data, cross-correlate it between three detector pairs (HL, HV, and LV), and coarse-grain the resulting spectra to a frequency resolution of 1/32 Hz. The cross-correlated data from each detector pair is then folded into one sidereal day. Finally, the folded cross-correlated data in the frequency domain are combined from different 192-second sidereal segments and detector pairs using Eqs. \eqref{eq:dirtymap} and \eqref{eq:fisher} to produce an estimate of GWB power $\mathcal{P}_\mu$ \cite{allen-romano,ain2015fast}. 

The sensitivities of cross-correlation based GWB searches are adversely affected by non-Gaussian features in the data. So we apply data quality cuts in time domain as well as in frequency domain to remove the non-Gaussian features associated with instrumental artifacts. Since the sensitivities of cross-correlation based searches are proportional to the square root of the total duration of the data analyzed and to the square root of total bandwidth used, these data quality cuts are a trade-off between the decrease in sensitivity due to non-Gaussian features and the decrease in sensitivity due to less time-frequency data being used.

In our analysis we use the same time domain cuts that were applied in the O3 isotropic analysis~\cite{isotropic2020} and only analyze data segments during which the detectors were in ``observing mode''~\cite{detchar2020}. We apply a nonstationarity cut to exclude data segments whose power spectral densities vary by more than 20\% relative to their neighboring segments. We remove the first two weeks of Hanford detector data due to nonstationarities around the calibration lines at $\sim 36$ Hz. Since we are interested in the GWB produced by events not explicitly detected by the LIGO-Virgo detector network, in addition to the above cuts, we also remove three segments ($3 \times 192$ seconds) worth of data around the published gravitational-wave events in the first half of O3~\cite{O3aCatalog}. Since there is no complete list of confirmed events for the second half of O3 yet, we remove times around the nonretracted gravitational-wave event candidates in the second half of O3~\cite{gwalerts}.

 During this observing run, the Livingston and Hanford detectors exhibited a large number of short-duration glitches \cite{detchar2020}. When left unchecked, these glitches induced non-Gaussian effects in the cross-correlation and autocorrelation power spectral density estimates and hence the nonstationarity data cuts employed vetoed a significant fraction of the viable data ($>$50\%). Since sensitivities of the cross-correlation based searches are proportional to the square root of the total duration of the data analyzed, these glitches significantly reduce the sensitivities of the searches. This prompted the development of a \emph{gating} procedure \cite{gatingDocument,y_stochasticGatingDocument} which excludes the glitches by applying an inverse Tukey window to Livingston and Hanford data at times when the root-mean-square value of the whitened strain channel in the 25-50 Hz band or 70-110 Hz band exceeds a certain threshold. Gating has proven effective: more data remains after the nonstationarity cuts, and the background power spectral density behaves as expected for uncorrelated Gaussian noise \cite{isotropic2020}. Furthermore, because of gating, the nonstationarity cuts only remove 10.7\%, 14.3\%,and 14.7\% of segments from HL, HV and LV baselines, respectively. Consequently, we analyzed 169 days of live time for the HL baseline, 146 days for the HV baseline, and 153 days for the LV baseline (which are longer than the 129 days of live time for the first two observing runs combined~\cite{abbott2019directional}).

For the analyses reported in this paper, we use the frequency band between 20 and 1726 Hz. In addition to the time-domain cuts, we also remove problematic frequencies from the analysis band. These frequencies are typically associated with known instrumental features such as calibration lines, power lines and their harmonics, hardware injections of continuous gravitational-wave signals, etc. \cite{o1_o2_lines_paper}. These frequencies are identified through coherence studies between detector strain data as well as data from auxiliary channels from the detector sites. In our analysis, we remove problematic frequencies identified for the O3 isotropic stochastic analysis \cite{isotropic2020}. 

\begin{figure}[htbp]
\centering
\includegraphics[width=0.95\columnwidth]{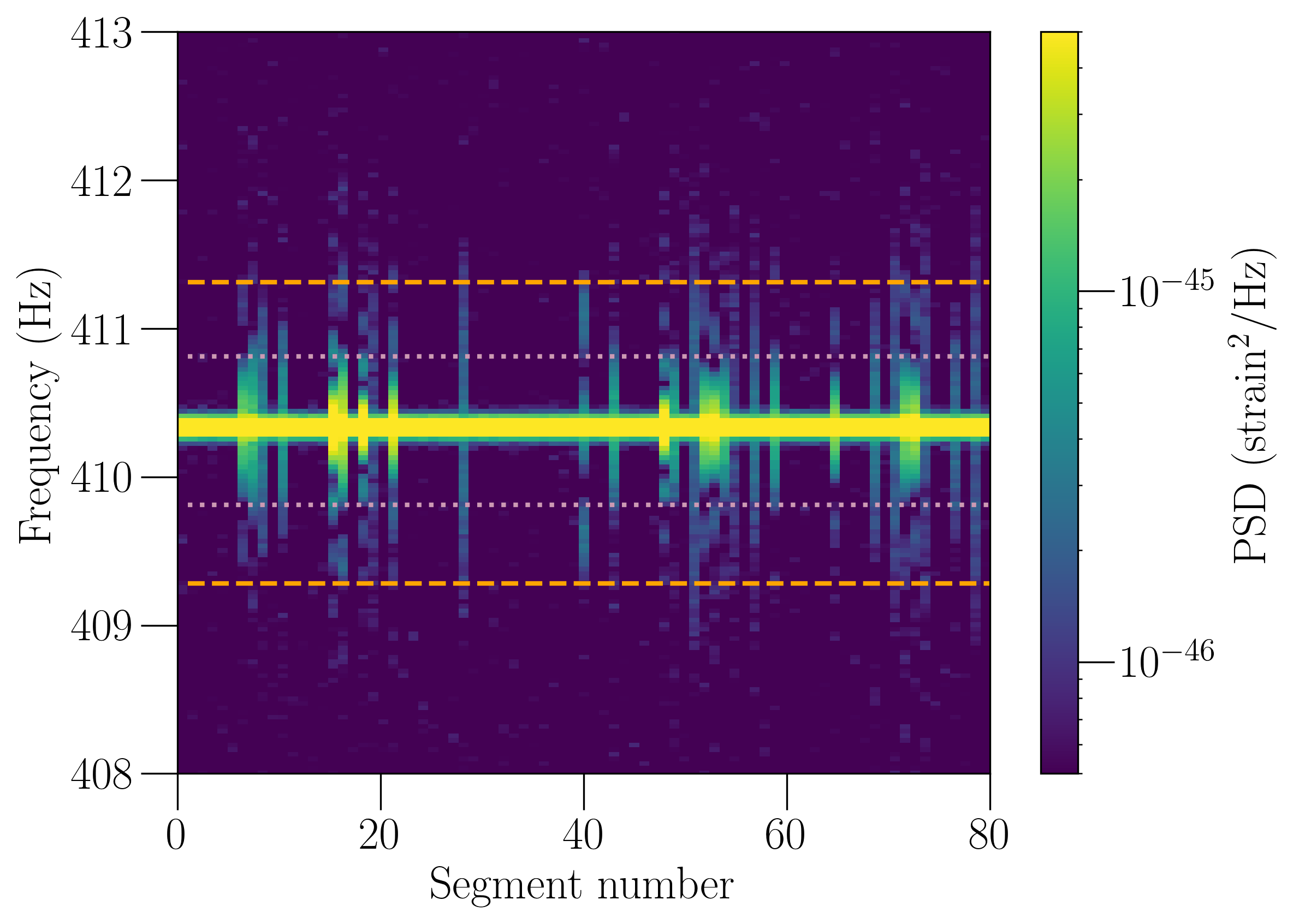}
\caption{Power spectral density spectrogram of the Hanford detector data around the 410.3 Hz calibration line using a short stretch of {\it gated} data. Each vertical column in the above plot corresponds to a power spectral density estimate using a 192 second long segment. The purple dotted lines show the region of the standard frequency notch around the calibration line and the orange dashed lines show the frequency region notched based on the nonstationarity level. We see that the latter removes a good portion of the nonstationarity region around the calibration line.}
\label{fig:nonstationarity_cut}
\end{figure}

Even though gating removes short-duration transients from the data, it introduces spectral artifacts around strong lines, such as calibration lines, in detector data that significantly affect the NBR analysis. These spectral artifacts behave similar to nonstationarities around those strong lines (for example as shown in Fig.~\ref{fig:nonstationarity_cut}). Hence we apply a threshold cut on the nonstationarity level in individual detector power spectral densities to remove these frequency regions of spectral artifacts. We remove frequencies when the standard deviation of the power spectral density at those frequencies exceeds the median power spectral density obtained from the entire run. The final list of frequencies notched in our current analysis can be found in Ref. \cite{dataDirectional2020}. We note here that the sensitivity loss due to these additional frequency notches from using gated-data is at the level of a few percent while the sensitivity loss due to not using gated-data is at the level of $>40$ \%. Hence we use gated-data with the additional frequency notching in our analysis. 
Removing all these frequencies cut approximately 14.8\%, 25.2\% and 21.9\% of usable bandwidth from the HL, HV and LV baselines respectively. We note here that although we use the frequency band between 20 and 1726 Hz, 99\% of sensitivity for broadband analyses comes from $\approx 20 - 300$ Hz band \cite{isotropic2020}.

\section{Results and Discussions} \label{sec:results}
\subsection{Broadband radiometer}
\label{sec:BBR_results}
\begin{table*}
    \begin{tabular}{ccc | cc ||cccc|| c | c }
    \multicolumn{5}{c}{\bf All-sky BBR Results}& \multicolumn{6}{c}{} \\
   \multicolumn{5}{c}{}& \multicolumn{4}{c}{Max SNR  (\% $p$-value)} & \multicolumn{2}{c}{Upper limit ranges $(10^{-8})$}\\
     \hline
     &$\alpha$&  & $\Ogw$ & $H(f)$ & HL(O3) &  HV(O3) & LV(O3) & O1+O2+O3 (HLV)  & O1+O2+O3 (HLV) & O1 + O2 (HL) \\
     \hline 
     &0&			& constant 			&  $\propto f^{-3}$        & 2.3  (66) & 3.4 (24) & 3.1 (51) & 2.6 (23) & 1.7 -- 7.6 & 4.4 -- 21  \\
     &$2/3$& 	& $\propto f^{2/3}$	&  $\propto f^{-7/3}$      & 2.5 (59) & 3.7 (14) & 3.1 (62) & 2.7 (24) &  0.85 --  4.1 & 2.3 -- 12 \\
     &3&			& $\propto f^{3}$	&  constant       		 & 3.7 (32) & 3.6 (47) & 4.1 (12) & 3.6 (20) & 0.013 -- 0.11 & 0.046 -- 0.32 \\
     \hline
   \end{tabular}
\caption{
The maximum SNR across all sky positions, its estimated $p$-value, and the range of the 95\% upper limits on gravitational-wave energy flux $F_{\alpha, \Theta}$  [${\text{erg cm}^{-2}\text{ s}^{-1} \text{ Hz}^{-1}}$] set by the BBR search for each baseline and for the three baselines combined using data from three LIGO observing runs and Virgo O3. The median improvement across the sky compared to limits from O2 analysis is a factor of 3.3 - 3.5, depending on $\alpha$. O1+O2 upper limits reported in the last column differ from the upper limits reported in \cite{abbott2019directional} for the reasons explained in the main text.}
\label{tab:alphas_BBR}
\end{table*}

\begin{figure*}[htbp!]
 \resizebox{\textwidth}{!}{
   \begin{tabular}{c}
    \boldmath{\hspace{1 cm} $\alpha = 0  \hspace{5 cm} \alpha=2/3  \hspace{5 cm} \alpha = 3$} \\ 
    {\bf SNR} \raisebox{-.65\height}{\includegraphics[width=\textwidth]{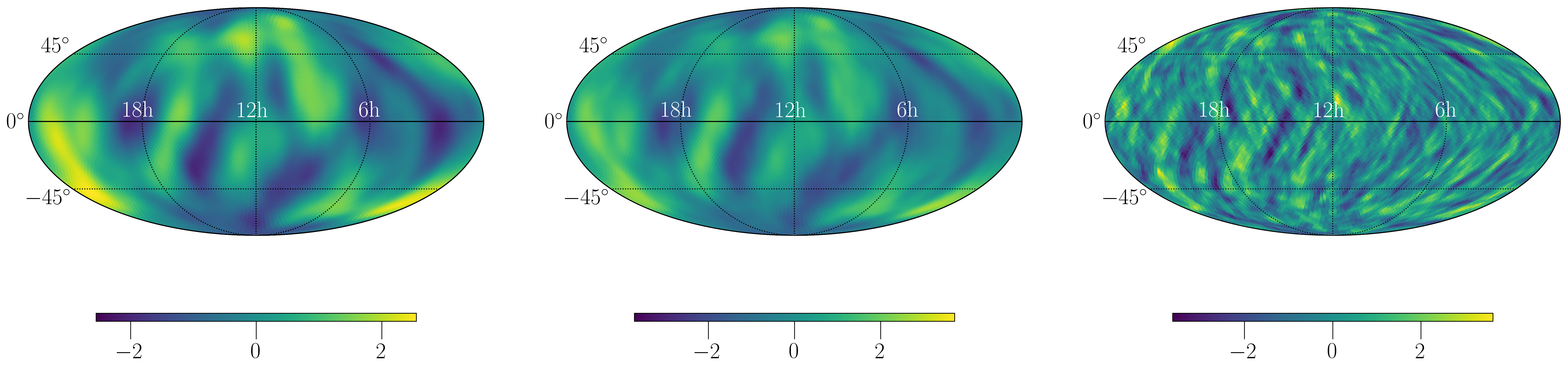}} \\
    {\bf UL} \raisebox{-.68\height}{\includegraphics[width=\textwidth]{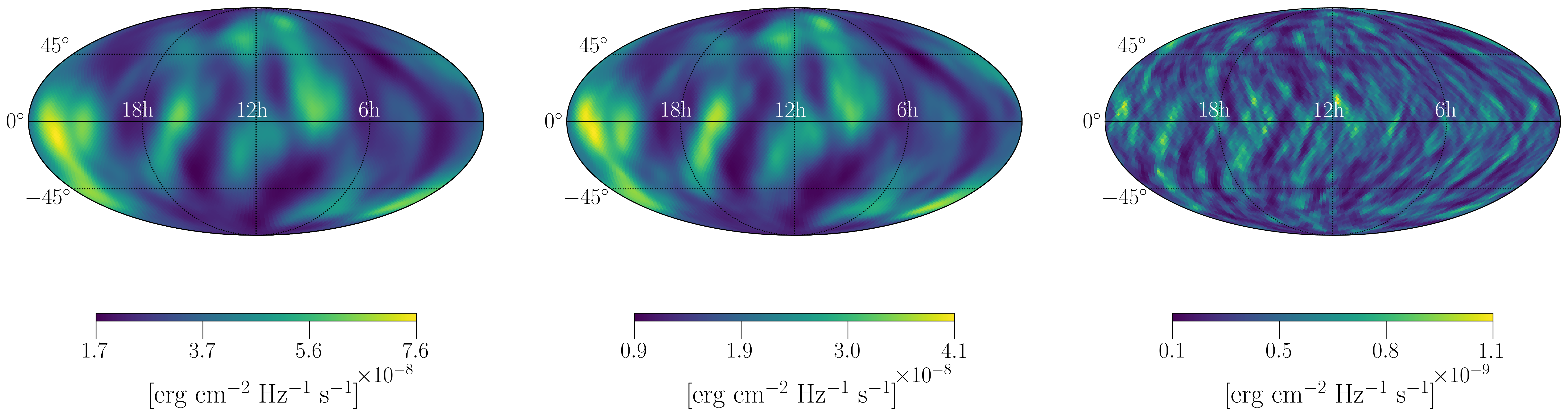}}
  \end{tabular}}
  \caption{Top row: SNR maps from a BBR search for pointlike sources. Bottom row: upper limit (UL) sky maps of the gravitational-wave energy flux. Both sets of maps, presented in equatorial coordinate system, are derived by combining all three LIGO observing runs and the Virgo O3 data. $\alpha = 0, \, 2/3$, and $3$ are represented from left to right.}
\label{fig:RADmaps_BBR}  
\end{figure*}

The sky maps obtained by combining data Eqs. (\ref{eq:combine_map}), (\ref{eq:combine_Fisher}) from LIGO-Virgo's past three observing runs (O1, O2 and O3) and from all three baselines HL, HV and LV (note that only O3 is used for HV and LV analysis) are shown in Fig. ~\ref{fig:RADmaps_BBR}, where each column refers to a different spectral index.
The top row shows the signal-to-noise ratio (SNR), which is the ratio of 
$\mathcal{P}_{\Theta}$ to $[\Gamma_{\Theta \Theta}]^{-1/2}$ in each sky direction. 

These SNR maps are consistent with Gaussian noise (see the $p$-values in Table ~\ref{tab:alphas_BBR}) and hence we place Bayesian upper limits, shown in the bottom row of Fig.~\ref{fig:RADmaps_BBR}, on the gravitational-wave energy flux from different sky directions. Due to the covariance 
between different pixels on the sky, the maximum SNR distribution is computed numerically by simulating many realizations of the dirty 
map {$X_{\nu}$} Eq. ({\ref{eq:combine_map}}) with the covariances described by the Fisher matrix {$\Gamma_{\mu\nu}$} Eq. ({\ref{eq:combine_Fisher}}). This maximum SNR distribution is then used to calculate the {$p$}-values for a given sky map with certain maximum SNR.

To evaluate the upper limits, we have used the techniques presented in ~\cite{Whelan:2012ur}, where a posteriori is built from the multivariate likelihood of the point estimate $\hat{\mathcal{P}}_{\Theta}$ after a marginalization over the calibration uncertainties. For all the analyses reported in this paper, we use amplitude calibration uncertainties of 7.0\% for Hanford, 6.4\% for Livingston and 5\% for Virgo data \cite{calibrationO3}. 

In contrast to the past BBR analysis, where a Cartesian grid was used to pixelate the sky, here we employ {\tt HEALPix} pixelization scheme with $n_{\rm side} = 32$, which implies $12n_{\rm{side}}^2 = 12288$ pixels, each with an area of $3$ $\rm{deg}^2$. The maximum SNR values observed in the sky maps for different $\alpha$, their associated $p$-values, and 95\% confidence upper limits on the gravitational-wave flux are reported in Table ~\ref{tab:alphas_BBR}. These limits improve upon the previous limits from O1+O2 data by a median factor (across the sky) of
$3.3 - 3.5$, depending on $\alpha$. We note here that the O1+O2 upper limits reported in the last column of Table ~\ref{tab:alphas_BBR} differ from those available in \cite{abbott2019directional}. This is because we found that the list of frequencies notched in the O2 analysis was not the optimal one and hence we regenerated the O1+O2 results by applying the appropriate frequency notching \cite{O2_notchlist}. The differences between the new and old O1+O2 upper limits are at the level of $\sim 5\%$.

Fig.~\ref{fig:sigma_RADmaps_BBR_appendix} in the Appendix shows sensitivity maps of individual baselines for different values of $\alpha$. From these plots we see that the sensitivity of the HL baseline is $\sim 3 - 10$ times better than that of the HV and LV baselines, depending on $\alpha$. Hence the final combined upper limit results are dominated by the HL baseline.  

\subsection{Spherical harmonics analysis}
\label{sphsection}
\begin{figure*}[htbp!]
  \begin{tabular}{c}
  \boldmath{\hspace{1 cm} $\alpha = 0  \hspace{5 cm} \alpha=2/3  \hspace{5 cm} \alpha = 3$} \\
  {\bf SNR} \raisebox{-.65\height}{\includegraphics[width=\textwidth]{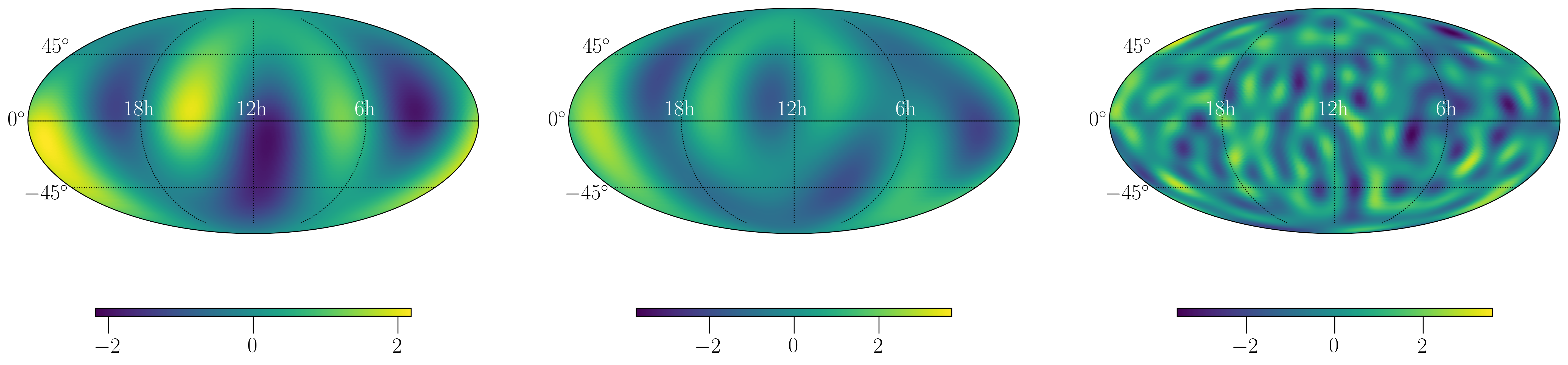}}\\
  {\bf UL} \raisebox{-.68\height}{\includegraphics[width=\textwidth]{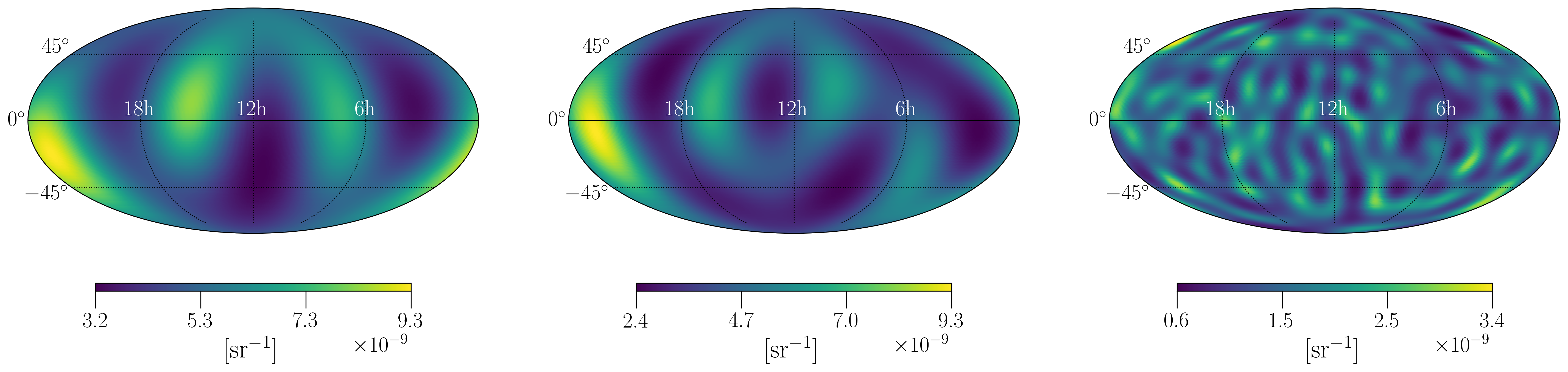}}\\
  \end{tabular}
  \caption{Top row: SNR maps from the SHD search for extended sources. Bottom row: sky maps representing 95\% upper limit on the normalized gravitational-wave energy density $\Omega_{\alpha}(\Theta) [{\rm sr^{-1}}]$. Both sets of maps, presented in equatorial coordinate system, are derived by combining all three observing runs of LIGO-Virgo data (Virgo was incorporated only for O3). $\alpha = 0, \, 2/3$, and $3$ are represented from left
  to right.}

\label{fig:SNRmaps_SHD}  
\end{figure*}

The sky maps obtained in the SHD analysis are presented in Fig.~\ref{fig:SNRmaps_SHD}, while a summary of the results is in Table~\ref{tab:alphas_SHD}. The maps presented in Fig.~\ref{fig:SNRmaps_SHD} are obtained by integrating over all available datasets (O1, O2 and O3) and running a combined analysis over the three baselines HL, HV, LV (O1 and O2 analyze only the HL baseline). However, sensitivity maps for the individual baselines are still useful to show how multiple baselines yield different anisotropy in its sensitivity, and are shown in Fig.~\ref{fig:sigma_map_baselines} in the Appendix.
In Fig.~\ref{fig:SNRmaps_SHD}, each column represents a different value of $\alpha$ and the top row shows the SNR maps while the bottom row shows 95\% confidence level upper limit maps.
According to the $p$-values in Table~\ref{tab:alphas_SHD}, the SNR sky
maps are consistent with Gaussian noise; hence we place upper limits on the normalized gravitational-wave energy density. Similar to BBR, 
the {$p$}-values in Table~{\ref{tab:alphas_SHD}} are calculated from the maximum SNR distribution computed numerically by simulating many realizations of the dirty map.
Table ~\ref{tab:alphas_SHD} also gives the range of upper limits in each sky map for combined data from LIGO-Virgo's three observing runs, as well as that from LIGO's O1+O2 analysis alone for comparison.
The Bayesian upper limits on the energy density spectrum have been derived based on posterior samples of $\hat{\mathcal{P}}_{\ell m}$ after marginalizing over the calibration uncertainties (see Ref.~\cite{Whelan:2012ur} for more details on how we treat calibration uncertainties). 

\begin{figure}[htbp]
\centering
\includegraphics[width=0.9\columnwidth]{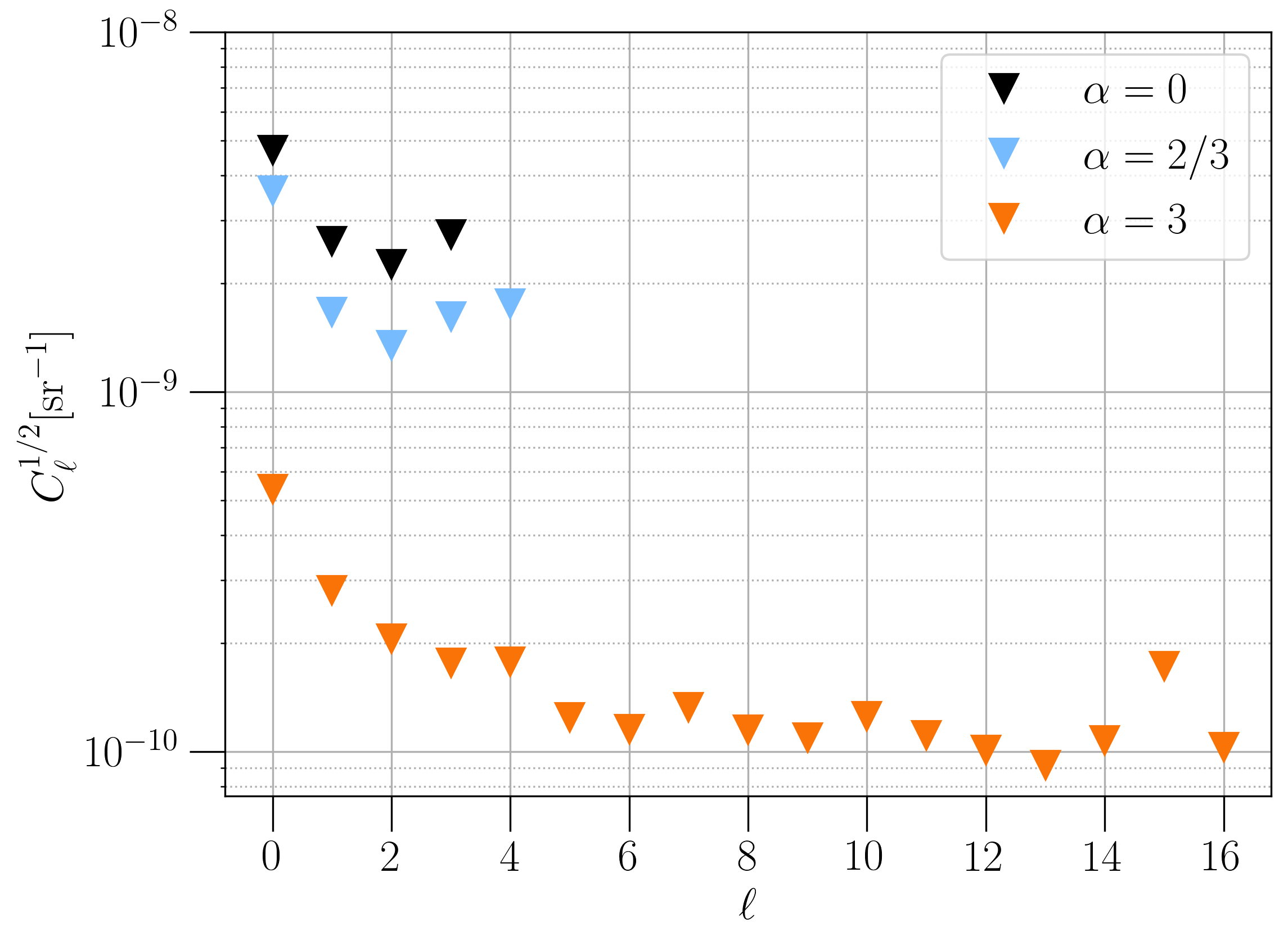}
\caption{$95\%$ upper limits on $C_\ell$ for different $\alpha$ using combined O1+O2+O3 data. }
\label{fig:Cl_limits}
\end{figure}

Additionally, in Fig.~\ref{fig:Cl_limits} we present the upper limits on $C_\ell^{1/2}$ at each angular scale $\ell$ for different signal models. 
The upper limits are improved by factors of $2.9 - 3.3$ with respect to the previous search \cite{abbott2019directional}. In contrast to $\Ogw(\Theta)$, the upper limits on $C_\ell$ are computed by constructing the Bayesian posteriors from the Monte Carlo sampling because the analytic expression for the probability distribution of $C_\ell$ is not trivial \cite{sph_methods}. Similarly, we marginalize the posteriors over calibration uncertainties.

\begin{figure}[htbp]
\centering
\includegraphics[width=0.9\columnwidth]{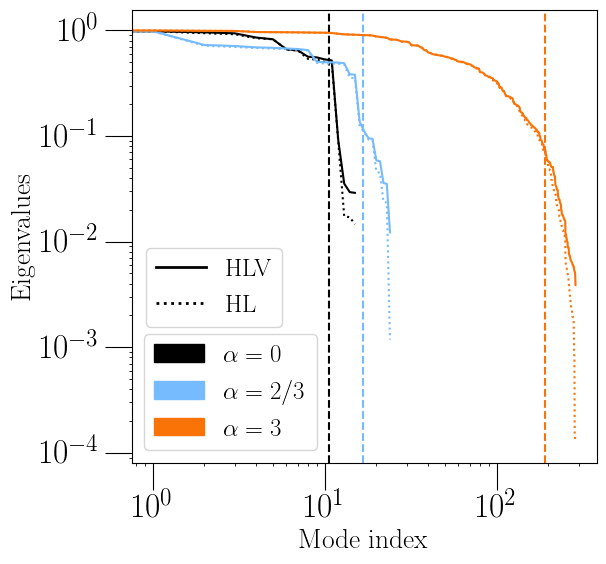}
\caption{Comparison between the Fisher matrix condition numbers for the HL and HLV networks for different values of $\alpha$. The vertical dashed lines mark the divisions between the two thirds of eigenvalues that are included in the analysis and one third that are excluded when inverting the Fisher matrix.}
\label{fig:fisher_condition}
\end{figure}

The impact of the new baselines on the SHD search may be quantified by monitoring the conditioning of the Fisher matrix, which is typically defined by the ratio of the largest to smallest eigenvalue of the matrix. The normalized eigenvalues of $\Gamma_{\mu\nu}$ for the single LIGO baseline (HL) and for the three-baseline configuration (HLV) are compared in Fig.~\ref{fig:fisher_condition}.   
The additional baselines have not had a significant effect on the eigenvalue distribution, particularly at $\alpha=0$ and $\alpha=2/3$, and hence we maintain the traditional regularization method of removing the lowest 1/3 of the eigenvalues~\cite{sph_methods}. We expect that this is because the sensitivity of the Virgo detector is not yet comparable to that of its LIGO counterparts. However, the new network has improved in the $\alpha=3$ case, where the smallest eigenvalue has increased by about two orders of magnitude. As the overall network sensitivity improves, the Fisher matrix will naturally regularize and higher modes will potentially be included in the reconstruction, enabling access to a higher resolution in the SHD search. This is in line with the projected results for multibaseline networks presented in Ref.~\cite{Renzini:2018vkx}. 

Below we consider the implications of our results for different astrophysical models. For
$\alpha=2/3$, the upper limit
found here for the corresponding $\ell$ modes is $C_\ell^{1/2} < 1.9 \times 10^{-9} \, \mathrm{sr}^{-1}$, whereas theoretical studies \cite{Jenkins:2018uac,Jenkins:2018kxc,Cusin:2018rsq} set 
$C_{\ell}^{1/2} \sim 10^{-12} \, \mathrm{sr}^{-1}$ for $1\leq \ell \leq 4$, assuming the normalized gravitational-wave energy density due to an isotropic GWB of compact binaries is $\sim 10^{-9}$~\cite{isotropic2020}. 
It is important to note that the finite sampling of the compact binary coalescences event rate leads to a 
spectrally-white shot noise term $(C_\ell^\mathrm{shot})^{1/2}\sim 10^{-10}\,\mathrm{sr}^{-1}$ which is orders of magnitude larger than the anticipated true astrophysical 
power spectrum~\cite{Jenkins:2019uzp}.
This term scales as $\propto1/\sqrt{T_\mathrm{obs}}$, where ${T_\mathrm{obs}}$ is the observation time, which is the same scaling expected for the upper limits set by the SHD search. 
Shot noise may therefore limit future SHD searches, if the SHD sensitivity improves faster than $1/\sqrt{T_\mathrm{obs}}$ due to the improved detector sensitivity or due to the increased number of detectors.
An optimal statistical method to estimate the true angular spectrum in the presence of shot noise was proposed in ~\cite{Jenkins:2019nks}.

For $\alpha=0$, we find the upper limit for the dipole ($\ell=1$) component to be $C_1^{1/2} < 2.6 \times 10^{-9} \,
\mathrm{sr}^{-1}$, whereas the theoretical study on Nambu-Goto strings based on model 3 in Ref.~\cite{Jenkins:2018lvb}, combined with the most up to date constraints on $G\mu$ using the isotropic component of the GWB \cite{O3_cosmicstring}, $G\mu\lesssim 4\times10^{-15}$, sets $C_1^{1/2}\lesssim 10^{-12}\mathrm{sr}^{-1}$. This
dipole moment is kinematically caused by the Earth's peculiar motion, and other $C_\ell$ modes resulting from the intrinsic anisotropy are expected to be many orders of magnitude smaller than the dipole moment.
For both choices of the power spectra ($\alpha = 0$ and $\alpha = 2/3$), we conclude that the predictions of the 
theoretical models are consistent with the search results presented here. 

\begin{table*}
    \begin{tabular}{ccc | cc ||cccc||c | c }
    \multicolumn{4}{c}{\bf SHD Results}\\
    \multicolumn{5}{c}{}& \multicolumn{4}{c}{Max SNR  (\% $p$-value)}&\multicolumn{2}{c}{Upper limit range $(10^{-9})$} \\
    \hline
     &$\alpha$&  & $\Ogw$ & $H(f)$ & HL(O3) &  HV(O3) & LV(O3) & O1+O2+O3 (HLV)  & O1+O2+O3 (HLV) & O1 + O2 (HL) \\
     \hline 
     &0& & constant & $\propto f^{-3}$ & 1.6 (78) & 2.1 (40) & 1.5 (83) & 2.2 (43) & 3.2--9.3  & 7.8--29 \\
     &$2/3$& & $\propto f^{2/3}$& $\propto f^{-7/3}$ & 3.0 (13) & 3.9 (0.98) & 1.9 (82) & 2.9 (18) & 2.4--9.3 &6.4--25 \\
     &3& & $\propto f^{3}$& constant & 3.9 (12) & 4.0 (10) & 3.9 (11) & 3.2 (60) & 0.57--3.4 & 1.9--11 \\
    \hline

    \end{tabular}
\caption{
We present the maximum SNR across all sky positions with its estimated $p$-value for the three separate baselines in the O3 observing as well as all three observing runs combined. We also present the range of the 95\% upper limits on the normalized gravitational-wave energy density $\Omega_{\alpha}(\Theta) [{\rm sr^{-1}}]$ after combining data from LIGO-Virgo's three observing runs. Note that for both the $p$-values and the upper limits, Virgo-related baselines are incorporated only for O3. The median improvement across the sky compared to limits set by the O1+O2 analysis is $2.9 - 3.3$ for the SHD search, depending on $\alpha$.}\label{tab:alphas_SHD}

\end{table*}


\subsection{Narrow band radiometer}
\begin{figure*}
  \begin{tabular}{ccc}
  \includegraphics[width=2.15in]{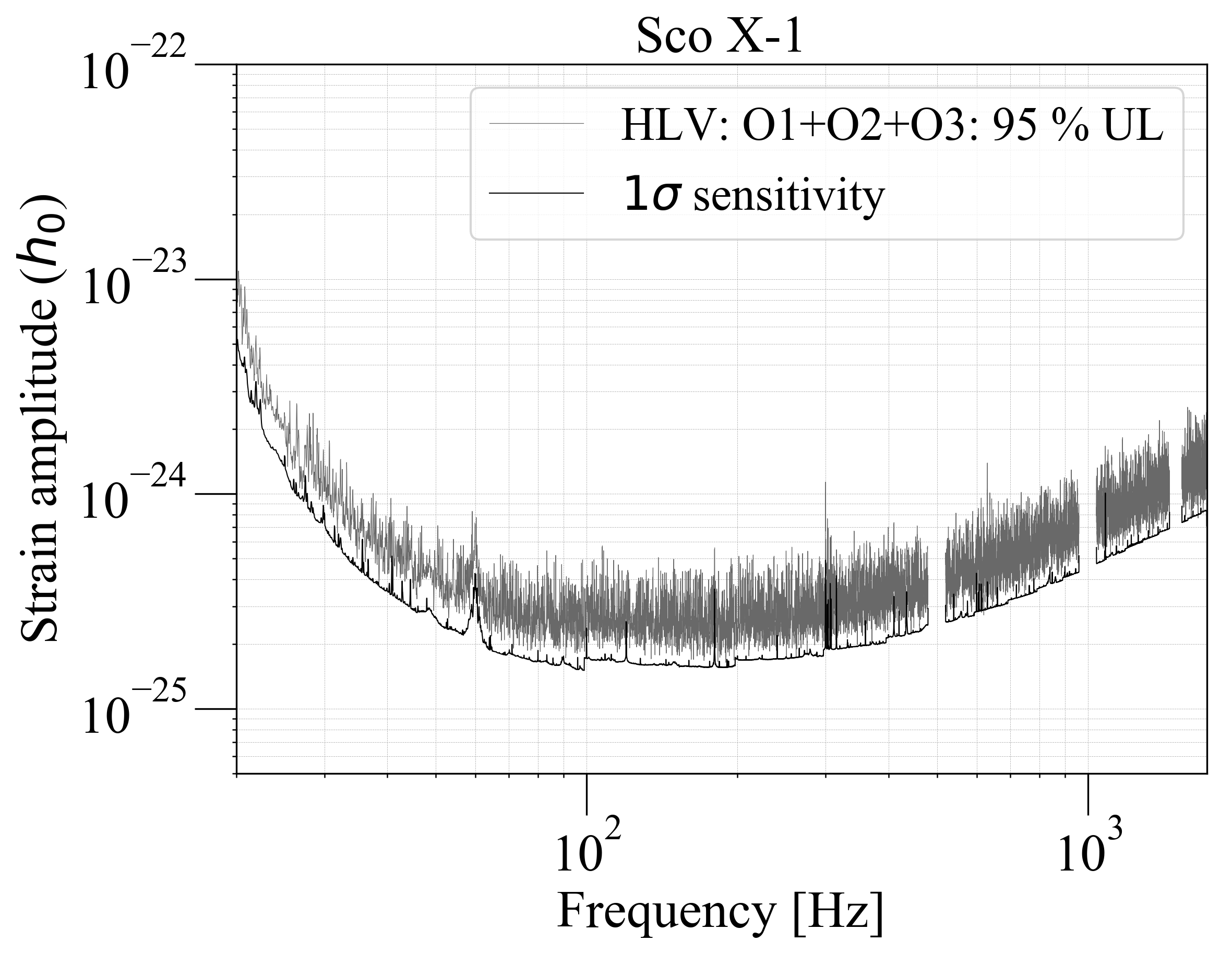} \hspace{.4cm} &
  \includegraphics[width=2.15in]{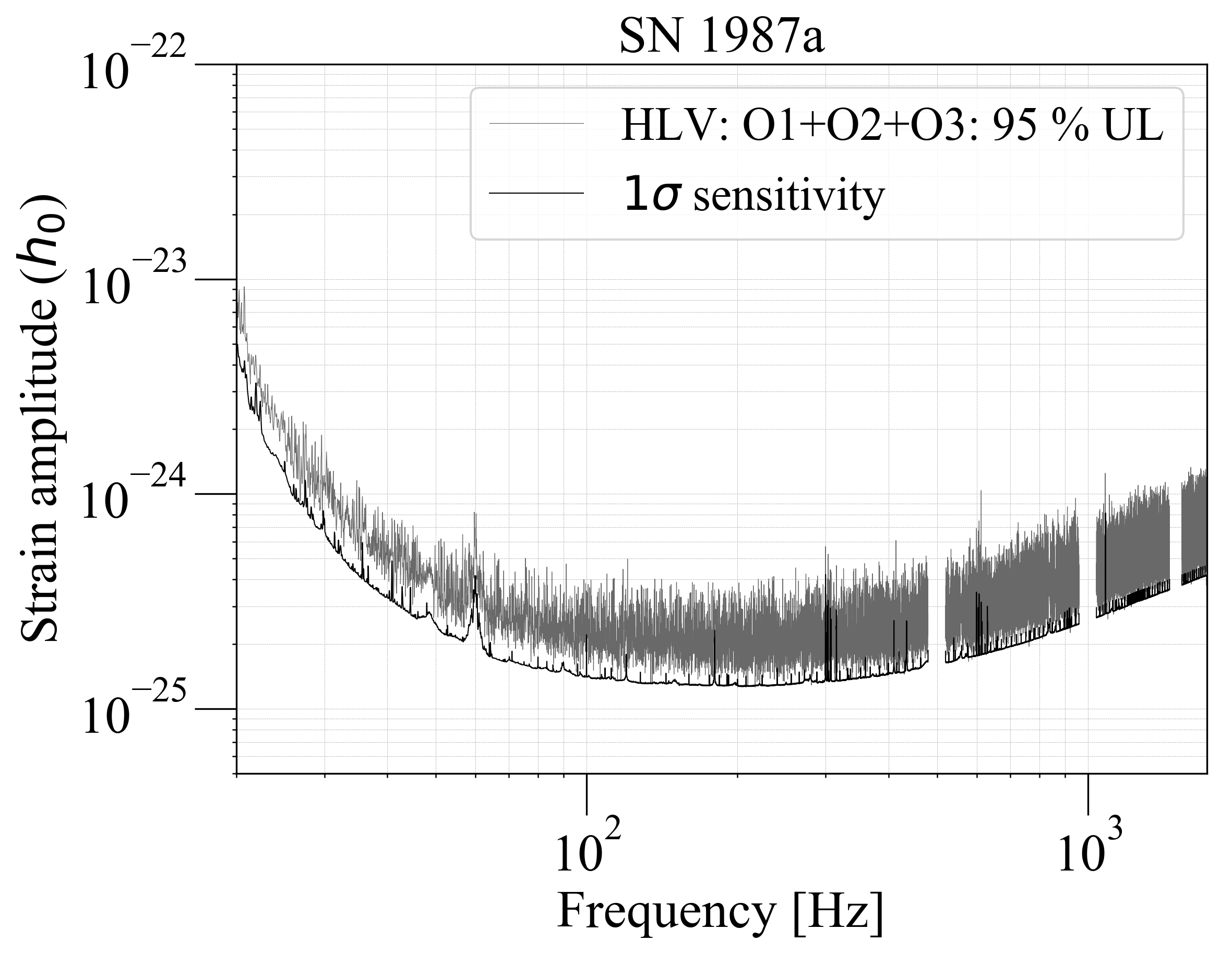} \hspace{.4cm} &
  \includegraphics[width=2.15in]{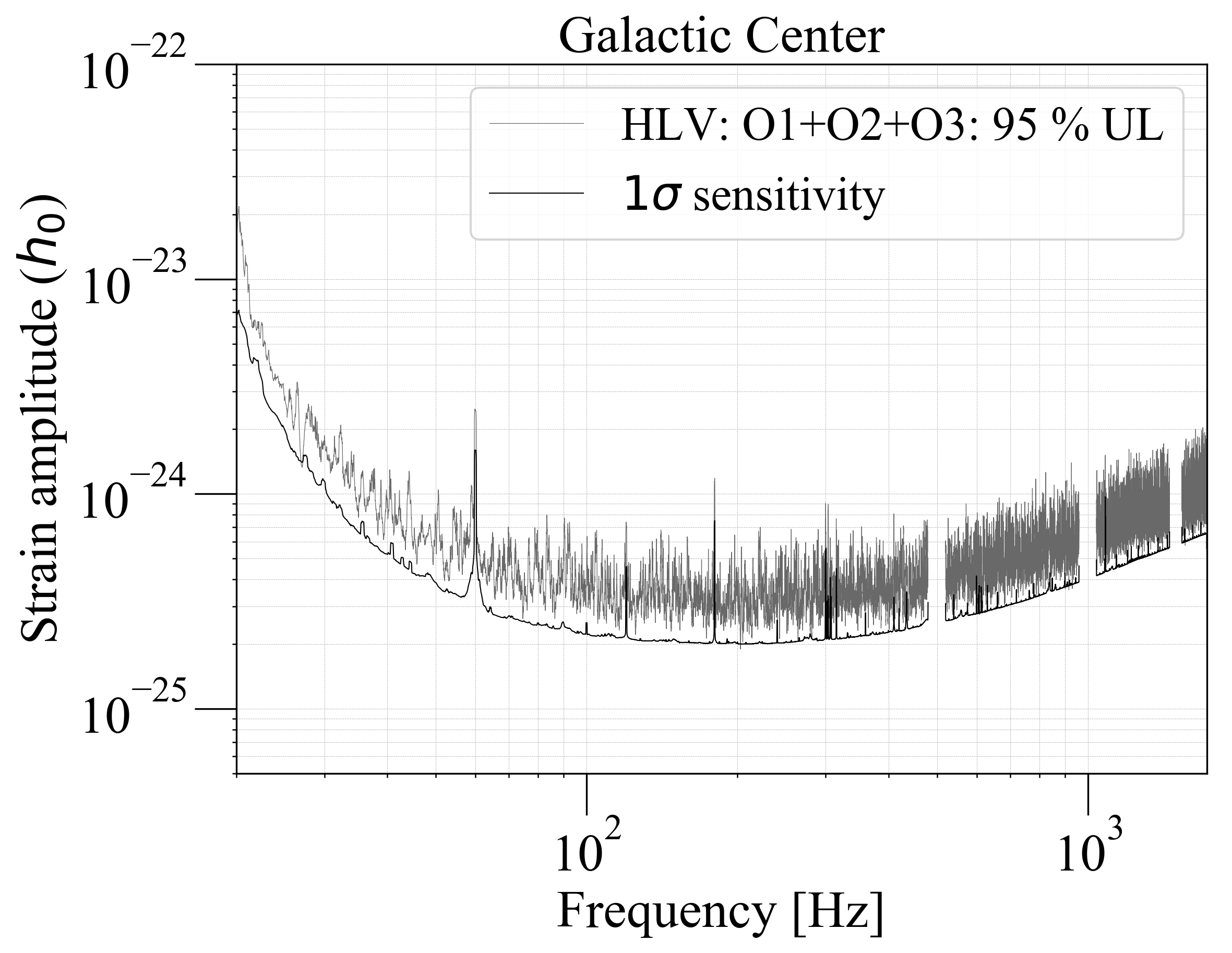} \hspace{.4cm} \\
  \end{tabular}
  \caption{
Upper limits on the dimensionless strain amplitude $h_0$, using the data from three observing runs of LIGO-Virgo detectors, at the 95\% confidence level for the narrow band radiometer search are indicated by the gray bands for Scorpius X-1 (left), SN 1987A (middle) and the Galactic Center (right). The dark line shows the 1$\sigma$ sensitivity of the search for each direction.  } 
\label{fig:NBR}
\end{figure*}

The gravitational-wave strain spectra obtained from the NBR search for each sky direction considered are shown in Fig.~\ref{fig:NBR}. For all three directions, we computed the SNR by combining the appropriately sized frequency bins across the three detectors. The maximum SNR across
the frequency band and an estimate of its significance are given in Table~\ref{tab:NBR} for each search direction. 
Our results are consistent with Gaussian noise in all three directions. We don't see any significant frequency outlier with $p$-value less 
than 1\%. Here the $p$-values are calculated from the maximum SNR distribution obtained by simulating many realizations of strain power consistent with Gaussian noise in each frequency bin and then combining the bins the same way as done in the actual analysis.

Since we do not find any compelling evidence for narrow-band gravitational waves, we set $95\%$ confidence limits on the peak strain amplitude $h_0$ (\footnotesize{$=\sqrt{\hat{\mathcal{H}}(f)}$} \normalsize) for each set of optimally combined frequency bins. When calculating this upper limit, we account for the Doppler modulation of the signal and marginalize over the inclination angle and polarization of the source. These limits, along with the $1\sigma$ sensitivity on $h_0$, are shown in Fig.~\ref{fig:NBR}. Since the limits fluctuate significantly due to the use of narrow frequency bins, we take a running median of them in a 1 Hz region around each frequency bin and report the best among these values as done in previous analyses \cite{abbott2019directional}.  
These limits correspond to an improvement by a factor of $\geq 2.0$ compared to limits from previous such analyses \cite{abbott2019directional}. The upper limits from individual baselines are shown in Fig.~\ref{fig:NBR_appendix} in the Appendix. 

\begin{table*}
  \begin{tabular}{ccc|ccc|cc}
   \multicolumn{5}{l}{\bf Narrowband Radiometer Results} \\
   \hline
   \multicolumn{2}{c}{Direction}&& Max SNR & $p$-value (\%) & Frequency (Hz) ($\pm~\unit[0.016]{Hz}$) & Best upper limit $(10^{-25})$ & Frequency band (Hz) \\
    \hline 
    &Scorpius X-1                    && 4.1  & 65.7 & 630.31   & 2.1 & $ 189.31 - 190.31$ \\
    &SN 1987A                   &&4.9  & 1.8 & 414.0  & 1.7 & $ 185.13 - 186.13$ \\
    &Galactic Center    && 4.1  & 62.3 & 927.25 & 2.1 & $ 202.56- 203.56$ \\ \hline
  \end{tabular}
  \caption{We show the maximum SNR, its estimated $p$-value, and the frequency bin of the maximum SNR for each search direction. We also give the best 95\% confidence level gravitational-wave strain upper limits achieved, and the corresponding frequency band, for all three sky locations. The best upper limits are taken as the median of the most sensitive 1 Hz band. All these results are derived from the three observing runs of LIGO-Virgo detectors.} \label{tab:NBR}
\end{table*}

It is meaningful to compare the upper limits in Fig. 6 with those derived in continuous-wave searches for neutron stars in past observing runs. Gravitational waves from Scorpius X-1 have been constrained using model-based cross correlation and hidden Markov Models using data from the first two Advanced LIGO/Virgo runs \cite{abbott2017upper, abbott2017search, Abbott:2019uwg, Zhang:2020rph}. The upper limits reported for Scorpius X-1 from continuous-wave searches \cite{abbott2017upper, Abbott:2019uwg, Zhang:2020rph} using LIGO/Virgo O1 and O2 data are comparable to or better than the limits we obtained in our analysis. The limits from continuous-wave searches are expected to further improve with LIGO/Virgo O3 data. The improvements in the modeled continuous-wave searches come at the expense of higher computational cost. Compared to the continuous-wave searches \cite{abbott2017upper, Abbott:2019uwg, Zhang:2020rph}, the unmodeled radiometer analysis reported in this paper is computationally inexpensive and also covers a larger frequency band (20-1726 Hz) than \cite{Abbott:2019uwg, Zhang:2020rph}. Regarding SN 1987A, a directed search has also been performed \cite{Sun:2016idj} using data from the second year of LIGO/Virgo’s fifth science run, which gave upper limits of about a factor of two worse than those presented here. However, searches on advanced detector data would surely improve this upper limit. Additionally, searches towards the Galactic Center for continuous waves have been run on data from LIGO/Virgo’s previous observing runs \cite{dergachev2019loosely, Piccinni:2019zub}, and have derived limits in a smaller frequency band that tend to be at least a factor of two better than those quoted here. The difference in limits is expected because the searches in \cite{dergachev2019loosely, Piccinni:2019zub} use much longer Fast Fourier Transform times that are specifically tuned to the frequency analyzed.

In the previous O2 NBR analysis reported in \cite{abbott2019directional}, an outlier with an SNR of 5.3 at a frequency of 36.06 Hz was found in the direction of SN 1987A. If this outlier were a true signal and consistent with an asymmetrically rotating neutron star slowly spinning down, we would expect to see it again in our O1+O2+O3 analysis with an even greater SNR because we have included the third observing run that is longer and more sensitive than the previous two runs. However, we do not find a similarly high SNR at that frequency and hence conclude that the outlier present in the previous run's data is not consistent with a persistent gravitational-wave signal.

\section{Conclusions}\label{sec:conclusions}

We do not find evidence for gravitational-wave signals in any of the three analyses using data from the three observing runs of Advanced LIGO and Virgo. Hence, we placed 95\% confidence level upper limits on the gravitational-wave energy density due to extended sources on the sky, on gravitational-wave energy flux from different directions on the sky, and on the median strain amplitude from possible sources in the directions of Scorpius X-1, the Galactic Center, and SN 1987A. These limits improve upon previous similar results by factors of $2.0 - 3.5$. We attribute this improvement partly to observing for twice as long as before, $\sim \sqrt{2}$, and partly to the improvement in the LIGO detector sensitivities. As mentioned in Sec.~\ref{sec:BBR_results}, the inclusion of the Virgo detector only marginally improves the upper limits due to its higher noise level compared to the LIGO detectors. However, we expect the Virgo detector to improve its noise performance in the next observing runs \cite{observingScene}. Furthermore, as noted in Sec.~\ref{sphsection}, the addition of Virgo detector to the detector network acts as a natural regularizer in the SHD analysis and would enable us to probe finer structures in the gravitational-wave sky maps. Currently we use flat, positive priors for the estimators {$\hat{\cal P}_\mu$} and in future analyses we plan to use more informative priors as done in Refs. \cite{Payne:2020pmc,taylor_pulsar_aniso,banagiri2021mapping}.
  
 As shown in \cite{isotropic2020}, the current GWB analyses are not affected by environmental effects, specifically magnetic correlation between the detectors. However as detector sensitivities improve, such environmental effects would become important and their effects on anisotropic GWB searches need to be studied.  Additionally, by taking advantage of folded data and new algorithms, we can perform an all-sky, all-frequency (ASAF) extension to the radiometer analysis for discovering persistent narrowband point sources \cite{borisanderic}. 
 
 As mentioned in Sec.~\ref{sphsection}, the current theoretical predictions for the anisotropies due to merger of compact objects, for example dipole component due to the Earth's peculiar motion, are more than an order magnitude below the upper limits presented in this paper. However with the planned enhancement of current generation of gravitational-wave detectors \cite{observingScene}, we might be able to measure these anisotropies. With the enhanced detector network, there is also possibility of detecting potential point sources of narrowband and broadband gravitational waves.   

Acknowledgments

The authors gratefully acknowledge the support of the United States
National Science Foundation (NSF) for the construction and operation of the
LIGO Laboratory and Advanced LIGO as well as the Science and Technology Facilities Council (STFC) of the
United Kingdom, the Max-Planck-Society (MPS), and the State of
Niedersachsen/Germany for support of the construction of Advanced LIGO 
and construction and operation of the GEO600 detector. 
Additional support for Advanced LIGO was provided by the Australian Research Council.
The authors gratefully acknowledge the Italian Istituto Nazionale di Fisica Nucleare (INFN),  
the French Centre National de la Recherche Scientifique (CNRS) and
the Netherlands Organization for Scientific Research, 
for the construction and operation of the Virgo detector
and the creation and support  of the EGO consortium. 
The authors also gratefully acknowledge research support from these agencies as well as by 
the Council of Scientific and Industrial Research of India, 
the Department of Science and Technology, India,
the Science \& Engineering Research Board (SERB), India,
the Ministry of Human Resource Development, India,
the Spanish Agencia Estatal de Investigaci\'on,
the Vicepresid\`encia i Conselleria d'Innovaci\'o, Recerca i Turisme and the Conselleria d'Educaci\'o i Universitat del Govern de les Illes Balears,
the Conselleria d'Innovaci\'o, Universitats, Ci\`encia i Societat Digital de la Generalitat Valenciana and
the CERCA Programme Generalitat de Catalunya, Spain,
the National Science Centre of Poland and the Foundation for Polish Science (FNP),
the Swiss National Science Foundation (SNSF),
the Russian Foundation for Basic Research, 
the Russian Science Foundation,
the European Commission,
the European Regional Development Funds (ERDF),
the Royal Society, 
the Scottish Funding Council, 
the Scottish Universities Physics Alliance, 
the Hungarian Scientific Research Fund (OTKA),
the French Lyon Institute of Origins (LIO),
the Belgian Fonds de la Recherche Scientifique (FRS-FNRS), 
Actions de Recherche Concertées (ARC) and
Fonds Wetenschappelijk Onderzoek – Vlaanderen (FWO), Belgium,
the Paris \^{I}le-de-France Region, 
the National Research, Development and Innovation Office Hungary (NKFIH), 
the National Research Foundation of Korea,
the Natural Science and Engineering Research Council Canada,
Canadian Foundation for Innovation (CFI),
the Brazilian Ministry of Science, Technology, and Innovations,
the International Center for Theoretical Physics South American Institute for Fundamental Research (ICTP-SAIFR), 
the Research Grants Council of Hong Kong,
the National Natural Science Foundation of China (NSFC),
the Leverhulme Trust, 
the Research Corporation, 
the Ministry of Science and Technology (MOST), Taiwan,
the United States Department of Energy,
and
the Kavli Foundation.
The authors gratefully acknowledge the support of the NSF, STFC, INFN and CNRS for provision of computational resources.
This work was supported by MEXT, JSPS Leading-edge Research Infrastructure Program, JSPS Grant-in-Aid for Specially Promoted Research 26000005, JSPS Grant-in-Aid for Scientific Research on Innovative Areas 2905: JP17H06358, JP17H06361 and JP17H06364, JSPS Core-to-Core Program A. Advanced Research Networks, JSPS Grant-in-Aid for Scientific Research (S) 17H06133, the joint research program of the Institute for Cosmic Ray Research, University of Tokyo, National Research Foundation (NRF) and Computing Infrastructure Project of KISTI-GSDC in Korea, Academia Sinica (AS), AS Grid Center (ASGC) and the Ministry of Science and Technology (MoST) in Taiwan under grants including AS-CDA-105-M06, Advanced Technology Center (ATC) of NAOJ, and Mechanical Engineering Center of KEK. 

We would also like to thank M. Alessandra Papa for providing useful comments that helped improve this paper.

The sky map plots have made use of healpy and HEALPix package \cite{healpy}. All plots have been prepared using Matplotlib \cite{matplotlib}.

{\it We would like to thank all of the essential workers who put their health at risk during the COVID-19 pandemic, without whom we would not have been able to complete this work.}

This document has been assigned the number LIGO-DCC-P2000500.

\appendix
\section*{Appendix: Individual baseline maps} \label{individualmaps}

Since this is the first time the Virgo detector has been used in the anisotropic GWB analysis, here we provide sensitivity maps for all the three baselines for comparison. However because of the relative low sensitivity of the Virgo detector compared to the LIGO detectors, the Hanford-Livingston baseline dominates the final results reported in the main part of the paper.

\begin{figure*}[htbp!]
  \boldmath{\hspace{1 cm} $\alpha = 0  \hspace{5 cm} \alpha=2/3  \hspace{5 cm} \alpha = 3$} \\
  \begin{tabular}{c}
    \includegraphics[width=\textwidth]{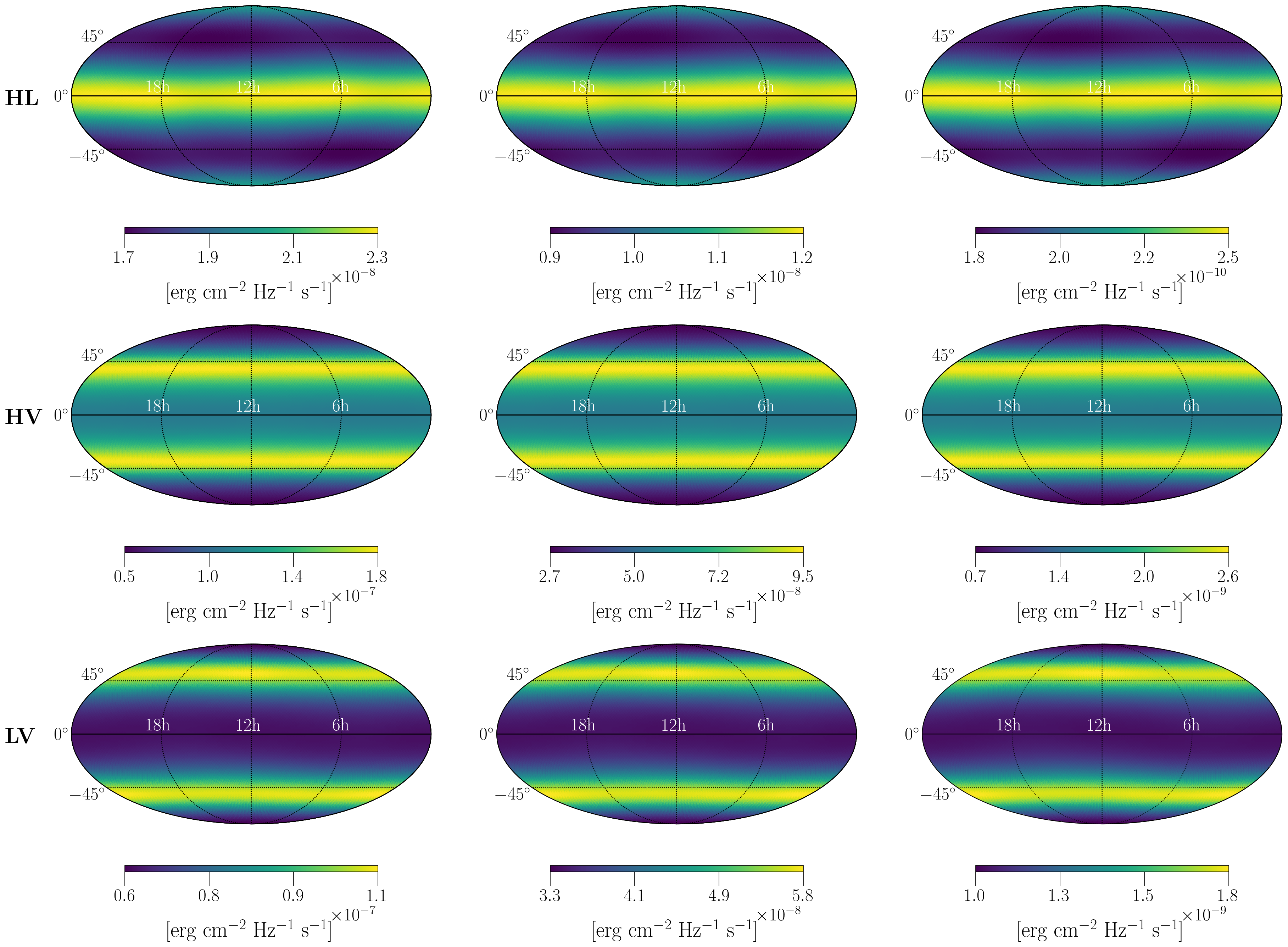}
  \end{tabular}
  \caption{Broadband radiometer maps illustrating search sensitivity for point like sources from O3 data only. Each row shows maps of the 1$\sigma$ sensitivity for HL, HV and LV baselines, from top to bottom, for three different power-law indices, $\alpha = 0, \,
2/3$ and $3$, from left to right.}  
\label{fig:sigma_RADmaps_BBR_appendix}  
\end{figure*}

\begin{figure*}[htbp!]
\boldmath{\hspace{1 cm} $\alpha = 0  \hspace{5 cm} \alpha=2/3  \hspace{5 cm} \alpha = 3$} \\
 \begin{tabular}{c}
	\includegraphics[width=\textwidth]{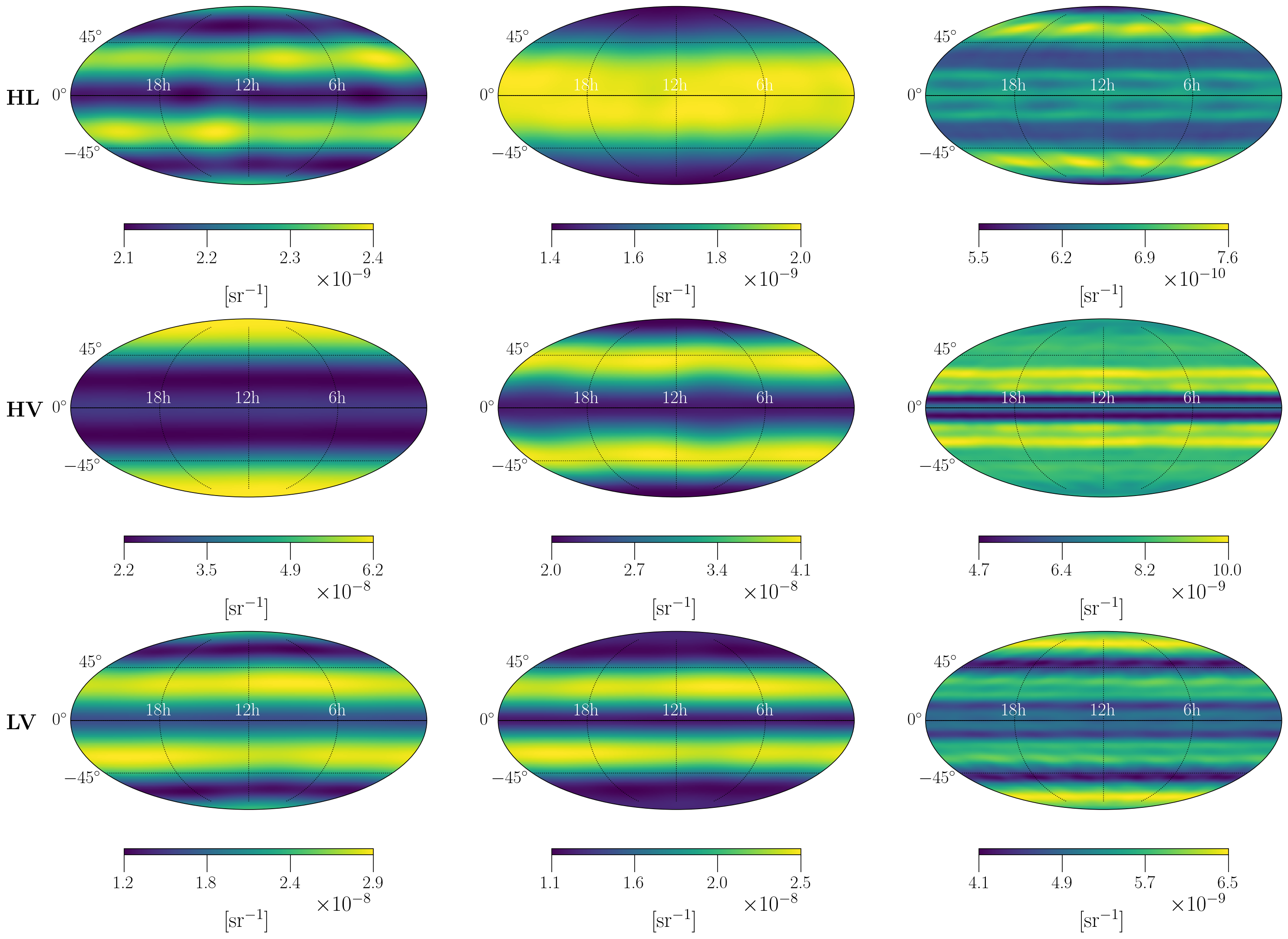}
\end{tabular}
	\caption{The spherical harmonics $1\sigma$ sensitivity maps produced from O3 illustrating
	a search for extended sources using each of HL, HV, LV baselines (top to
	bottom rows respectively). Three different power law indices, $\alpha =
	0, \, 2/3$ and $3$, are represented by columns from left to right.}
	\label{fig:sigma_map_baselines}
\end{figure*}

\begin{figure*}
 \begin{tabular}{ccc}
 \includegraphics[width=2.15in]{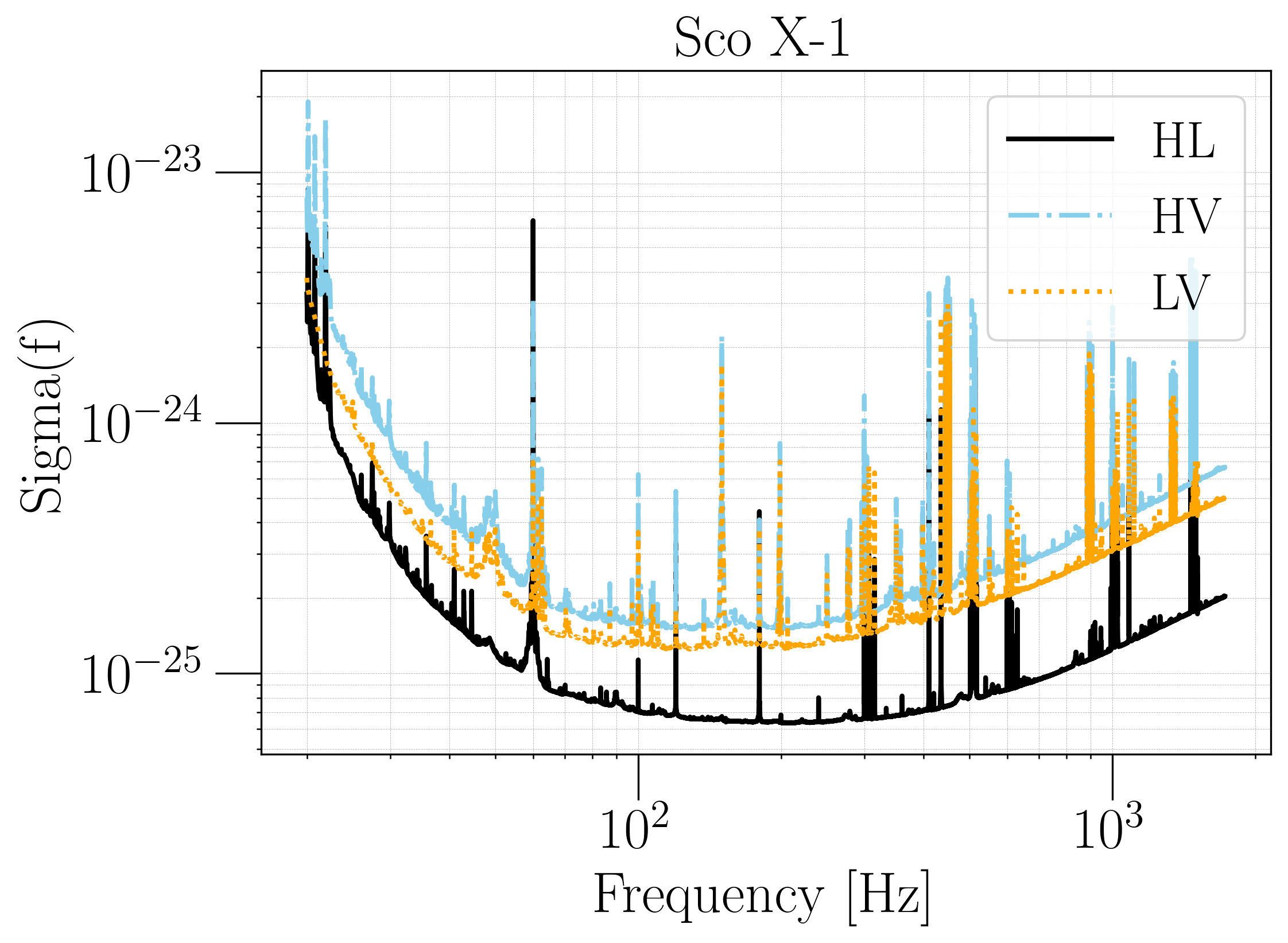} \hspace{.4cm} &
 \includegraphics[width=2.15in]{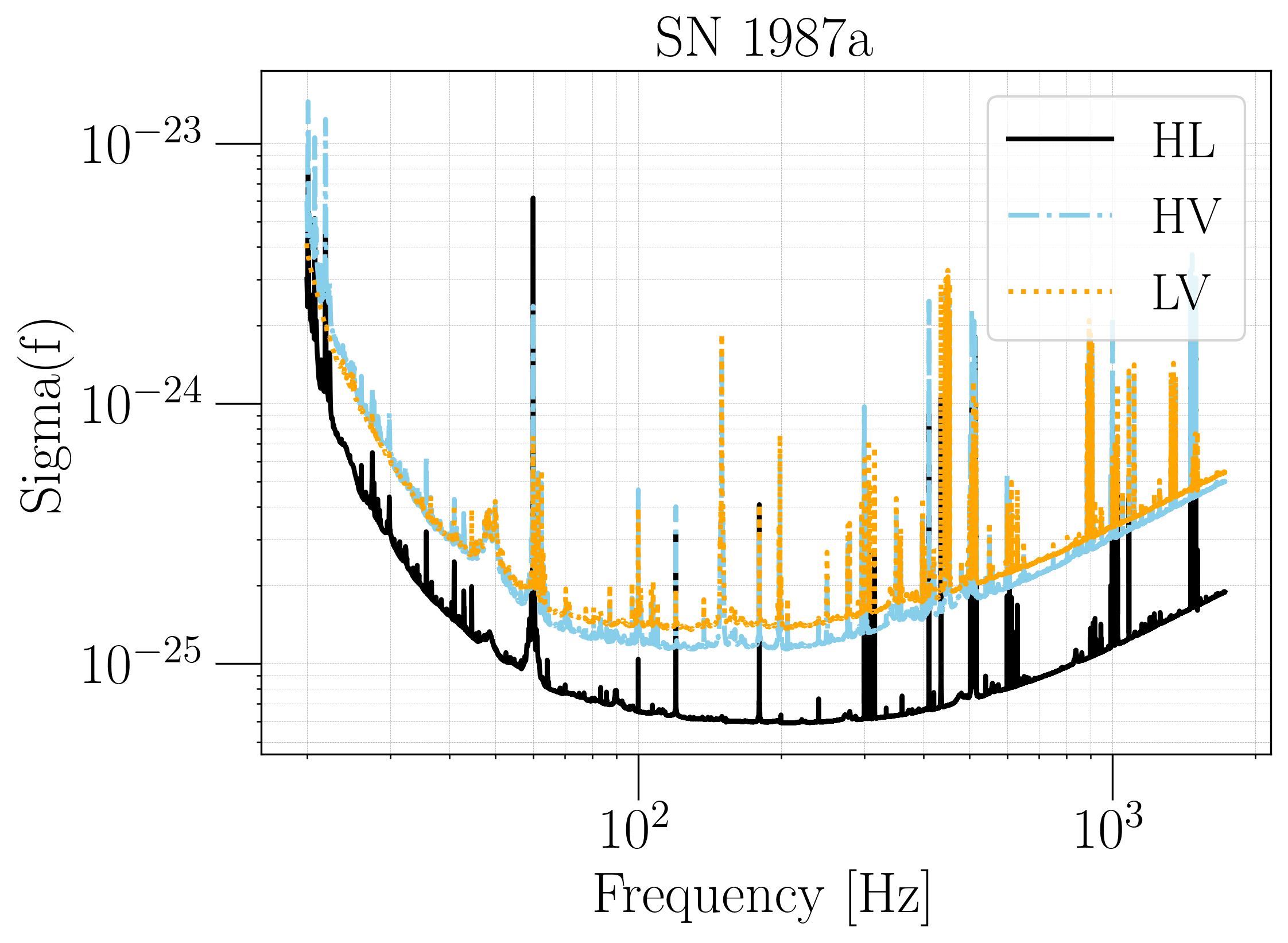} \hspace{.4cm} &
 \includegraphics[width=2.15in]{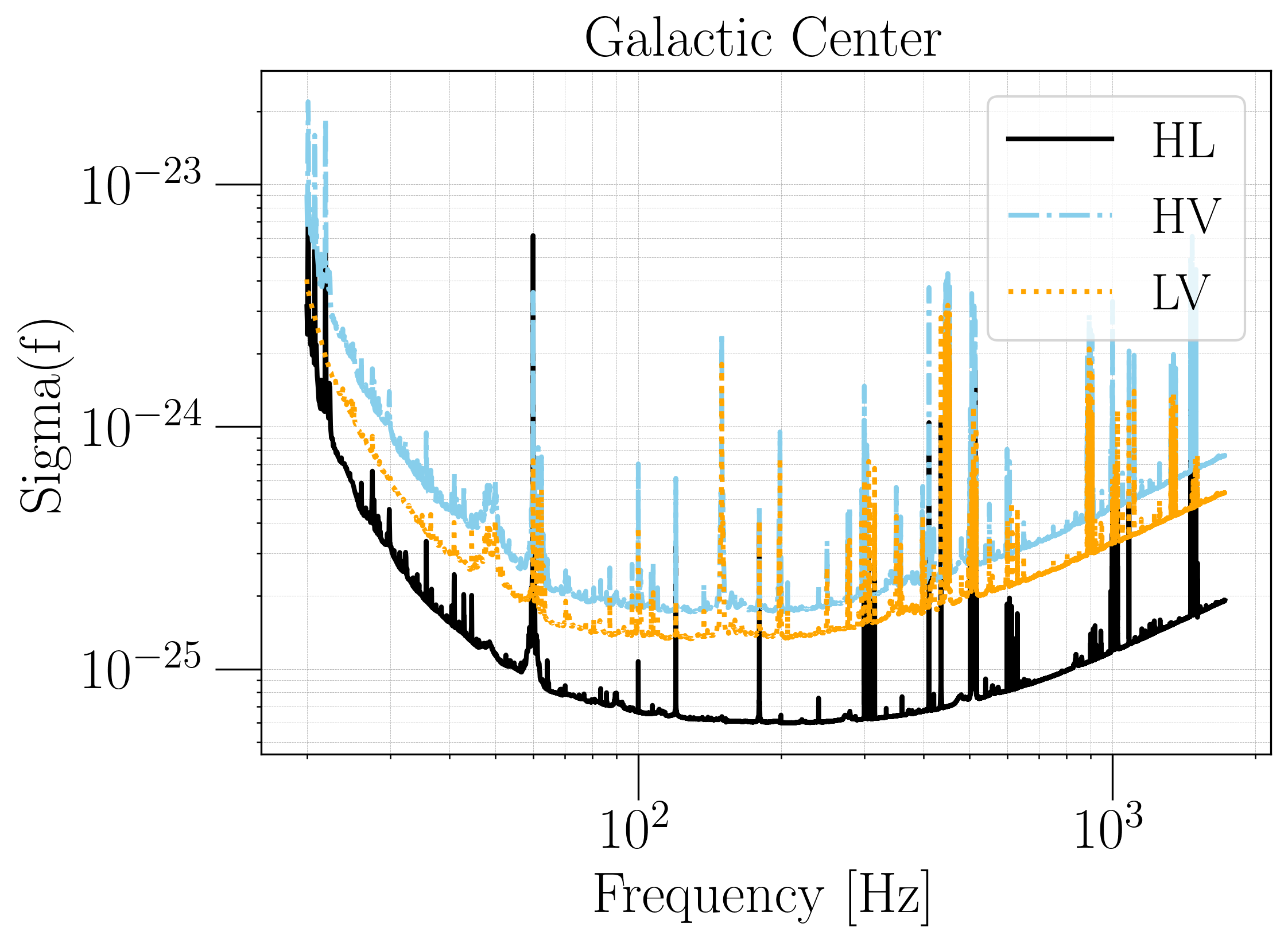} \hspace{.4cm} \\
 \end{tabular}
 \caption{The uncertainty associated with the NBR search estimator is shown. The uncertainty on the estimated gravitational-wave strain as a function of frequency is plotted for three directions, Scorpius X-1, SN 1987 and Galactic Center from left to right using O3 data for all the three baselines. HL baselines are indicated by the black lines, HV with sky blue lines and LV using orange.}
\label{fig:NBR_appendix}
\end{figure*}

\bibliographystyle{apsrev4-1}
\bibliography{refs}

\iftoggle{endauthorlist}{
  \let\author\myauthor
  \let\affiliation\myaffiliation
  \let\maketitle\mymaketitle
  \title{The LIGO Scientific Collaboration, Virgo Collaboration, and KAGRA Collaboration  }
  \pacs{}

%
\author{R.~Abbott}
\affiliation{LIGO Laboratory, California Institute of Technology, Pasadena, CA 91125, USA}
\author{T.~D.~Abbott}
\affiliation{Louisiana State University, Baton Rouge, LA 70803, USA}
\author{S.~Abraham}
\affiliation{Inter-University Centre for Astronomy and Astrophysics, Pune 411007, India}
\author{F.~Acernese}
\affiliation{Dipartimento di Farmacia, Universit\`a di Salerno, I-84084 Fisciano, Salerno, Italy  }
\affiliation{INFN, Sezione di Napoli, Complesso Universitario di Monte S.Angelo, I-80126 Napoli, Italy  }
\author{K.~Ackley}
\affiliation{OzGrav, School of Physics \& Astronomy, Monash University, Clayton 3800, Victoria, Australia}
\author{A.~Adams}
\affiliation{Christopher Newport University, Newport News, VA 23606, USA}
\author{C.~Adams}
\affiliation{LIGO Livingston Observatory, Livingston, LA 70754, USA}
\author{R.~X.~Adhikari}
\affiliation{LIGO Laboratory, California Institute of Technology, Pasadena, CA 91125, USA}
\author{V.~B.~Adya}
\affiliation{OzGrav, Australian National University, Canberra, Australian Capital Territory 0200, Australia}
\author{C.~Affeldt}
\affiliation{Max Planck Institute for Gravitational Physics (Albert Einstein Institute), D-30167 Hannover, Germany}
\affiliation{Leibniz Universit\"at Hannover, D-30167 Hannover, Germany}
% ...

\collaboration{The LIGO Scientific Collaboration, the Virgo Collaboration, and the KAGRA Collaboration}
%
%

  \newpage
  \maketitle
}

\end{document}